\documentclass[prd,nofootinbib,showpacs,superscriptaddress]{revtex4}

\usepackage{amsmath,amsfonts,amssymb,graphicx,subfigure}
\usepackage{hyperref}

\newcommand{\nysc}{{y\!\!\!\!-}}
\newcommand{\rmd}{\mathrm{d}}
\newcommand{\notp}[1]{\not\!{#1}}
\newcommand{\eq}[1]{\begin{equation} #1 \end{equation}} %

\begin{document}
\title{Pair production with neutrinos in an intense background
magnetic field}

\author{Duane A. Dicus}
\affiliation{Department of Physics, University of Texas at Austin,
Austin, Texas 78712}
\email{dicus@physics.utexas.edu}

\author{Wayne W. Repko}
\affiliation{Department of Physics and Astronomy, Michigan State
University, East Lansing, MI 48824}
\email{repko@pa.msu.edu}
\author{Todd M. Tinsley}
\affiliation{Physics and Astronomy Department, Rice University,
Houston, Texas 77005}
\email{tinsley@rice.edu}

\date{\today}

\begin{abstract}
We present a detailed calculation of the electron-positron
production rate using neutrinos in an intense background magnetic
field. The computation is done for the process $\nu \to \nu e
\bar{e}$ (where $\nu$ can be $\nu_e$, $\nu_\mu$, or $\nu_\tau$)
within the framework of the Standard Model. Results are given for
various combinations of Landau-levels over a range of possible
incoming neutrino energies and magnetic field strengths.
\end{abstract}


\maketitle

\section{\label{sec:intro}Introduction}

Neutrino interactions are of great importance in astrophysics
because of their capacity to serve as mediators for the transport
and loss of energy. Their low mass and weak couplings make neutrinos
ideal candidates for this role. Therefore, the rates of neutrino
interactions are integral in the evolution of all stars,
particularly the collapse and subsequent explosion of supernovae,
where the overwhelming majority of gravitational energy lost is
radiated away in the form of neutrinos.

Neutrinos have held a prominent place in models of stellar collapse
ever since Gamow and Schoenberg suggested their role in
1941~\cite{Gamow:1941a}. While supernova models have progressed a
great deal in the last 65 years, the precise mechanism for explosion
is still uncertain.  A common feature, however, among all models is
the sensitivity to neutrino transport.  Neutrino processes once
thought to be negligible now become relevant, and this has inspired
many authors to calculate rates for neutrino interactions beyond
that of the fundamental ``Urca'' processes
\begin{eqnarray*}
p\ e &\to& n\ \nu_e \\
n &\to& p\ e\ \bar{\nu}_e\,.
\end{eqnarray*}
Recent examples include neutrino-electron scattering,
neutrino-nucleus inelastic scattering, and electron-positron pair
annihilation~\cite{Bruenn:1991a,Mezzacappa:1993a}.  Furthermore, the
large magnetic field strengths associated with supernovae
($10^{12}$--$10^{17}$~G) are likely to cause significant changes in
the behavior of neutrino transport.

While the the electromagnetic field does not couple to the Standard
Model neutrino, it \emph{does} affect neutrino physics by altering
the behavior of any charged particles, real or virtual, with which
the neutrino may interact.  A number of authors have considered such
effects on Urca-type
processes~\cite{Dorofeev:1985az,Baiko:1998jq,Gvozdev:1999md,Arras:1998mv}
and on neutrino absorption by nucleons (and its reversed
processes)~\cite{Duan:2005fc,Bhattacharya:2002qf,Bhattacharya:2004nj}.
Furthermore, Bhattacharya and Pal have prepared a very nice review
of other processes involving neutrinos that are affected by the
presence of a magnetic field~\cite{Bhattacharya:2002aj}.

The problem of interest in this work is the production of
electron-positron pairs with neutrinos in an intense magnetic field
\begin{equation}\label{nu2eq:process}
\nu\ \to\ \nu\ e\ \bar{e}\,.
\end{equation}
Normally this process is kinematically forbidden,
but the presence of the magnetic field changes the energy balance of
the process, thereby permitting the interaction.

\begin{figure}
\subfigure[Neutral current
reaction\label{fig:zfig}]{\includegraphics[width=150pt]{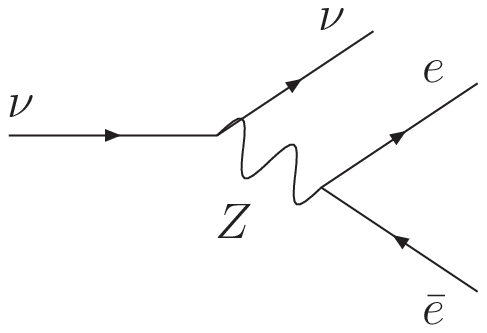}}
\qquad \subfigure[Charged current
reaction\label{fig:wfig}]{\includegraphics[width=150pt]{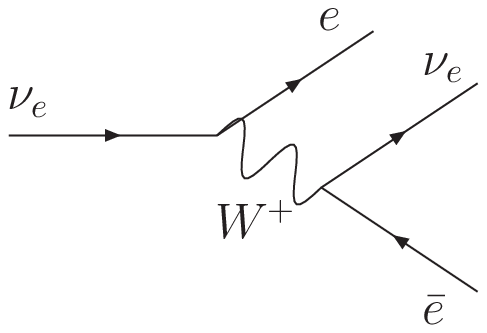}}
\caption{Possible diagrams considered for the process $\nu\ \to\ \nu
e \bar{e}$.  Both diagrams contribute for electron-type neutrinos,
but only the neutral current reaction (FIG.~\ref{fig:zfig})
contributes for $\nu_\mu$ and $\nu_\tau$.\label{fig:zwfigs}}
\end{figure}

Stimulation of this process with high-intensity laser fields has
been shown to have an unacceptably low rate of
production~\cite{Tinsley:2004pe}, but such an interaction could have
important consequences in astrophysical phenomena where large
magnetic field strengths exist. The process would most likely serve
to transfer energy in core-collapse
supernovae~\cite{Thompson:1993a}. However, Gvozdev \textit{et al.}\
have proposed that its role in magnetars could even help to explain
observed gamma-ray bursts~\cite{Gvozdev:1997bs}. The interest in
this reaction has led to a previous treatment in the
literature~\cite{Kuznetsov:1996vy}, but those authors present
results for two special limiting cases: (1) when the generalized
magnetic field strength $eB$ is greater than the square of the
initial neutrino energy $E^2$, and (2) when the square of the
initial neutrino energy $E^2$ is much greater than the generalized
magnetic field strength $eB$.  In both cases the incoming neutrino
energy $E$ is much greater than the electron's rest energy $m_e$. In
this paper we present a more complete calculation of the production
rate as mediated by the neutral and the charged-current processes
(FIG.~\ref{fig:zwfigs}).  We 
present the results of the calculation for varying Landau levels,
neutrino energies, and magnetic field strengths.  A comparison with
the approximate method is also discussed.

\section{Field operator solutions\label{sec:derive}}
As we have pointed out in section~\ref{sec:intro}, the standard
model neutrino can only be affected by the electromagnetic field
through its interactions with charged particles. This means that for
the process $\nu \to \nu e \bar{e}$ the Dirac field solution for the
final state electron and positron must change relative to their
free-field solutions.  The magnetic field will also change the form
of the $W$-boson's field solution which can mediate the process when
electron neutrinos are considered.  However, in our analysis we take
the limit that the momentum transfer for this reaction is much less
than the mass of the $W$-boson ($Q^2 \ll m_W^2$) and ignore any
effects the magnetic field may have on this charged boson.  Thus, in
this section we review the results of our derivation of the Dirac
field operator solutions for the electron and positron.  We closely
follow the conventions used by Bhattacharya and Pal and refer the
reader to their work~\cite{Bhattacharya:2002qf} for a detailed
derivation. The reader who is familiar with these solutions may wish
to begin with section~\ref{sec:prodrate} where we calculate the
production rate.

We choose our magnetic field to lie along the positive $z$-axis
\eq{\label{eq:B} \vec{B} = B_0 \hat{k}} which allows us some freedom
in the choice of vector potential $A(x)$.  We make the choice
\eq{\label{eq:A} A^\mu(x) = \left(0,-yB,0,0 \right)} both for its
simplicity and its agreement with the choice found in
reference~\cite{Bhattacharya:2002qf}.  This choice in vector
potential leads us to assume that all of the $y$ space-time
coordinate dependence is within the spinors.  The absence of any $y$
dependence in, for instance, the phase leads us to define a notation
such that \eq{\label{eq:ny} \nysc^\mu = \left(t, x, 0 , z\right)\,}
and
\eq{\label{eq:nvy} \vec{V}_{\nysc} = \left(V_x, 0, V_z\right)\,,} 
where $\vec{V}$ is any 3-vector.

\subsection{Electron field operator\label{ssec:efo}} %
Solving the Dirac equation for our choice of vector potential
(Eq.~\ref{eq:A}) results in the following electron field operator %
\eq{\label{eq:efieldop} \psi_e(x) = \sum_{n=0}^{\infty} \sum_{s=\pm}
\int\frac{\rmd^2 \vec{p}_{\nysc}}{(2\pi)^2} \sqrt{\frac{E_n + m_e}{2
E_n}} \left[ u^s(\vec{p}_{\nysc}, n, y)\, e^{- i p \cdot \nysc}\,
\hat{a}^s_{e\,\vec{p}_{\nysc}, n} + v^s(\vec{p}_{\nysc}, n, y)\,
e^{+ i p \cdot \nysc}\, {\hat{b}^{s\,
\dag}_{e\,\vec{p}_{\nysc}, n}} \right]\,,} %
where the creation and annihilation operators obey the following
anti-commutation relations %
\eq{\label{eq:eanticomm} \left\{ \hat{a}^s_{e\,\vec{p}_{\nysc}, n},
\hat{a}^{s'\, \dag}_{e\,{\vec{p}}'_{\nysc}, n'}\right\} = \left\{
\hat{b}^s_{e\,\vec{p}_{\nysc}, n}, \hat{b}^{s'\,
\dag}_{e\,\vec{p}'_{\nysc}, n'}\right\} = (2\pi)^2\, \delta^{ss'}\,
\delta_{nn'}\, \delta^2(\vec{p}_{\nysc} - \vec{p}_{\nysc}\,\!')\,.}  %

In Eq.~\ref{eq:efieldop} we sum over all possible spins $s$ and all
Landau levels $n$ where $E_n$ is the energy of fermion occupying the $n^\mathrm{th}$ Landau level %
\eq{\label{eq:En} E_n = \sqrt{p_z^2 + m_e^2 + 2 n e B}\,, \qquad n
\geq 0\,.}

The Dirac bi-spinors are
\begin{subequations}\label{eq:bispinors}
\eq{\label{eq:uspinor} u^+(\vec{p}_{\nysc}, n, y) =  \left(
\begin{array}{c} I_{n-1}(\xi_-)
\\ 0 \\ \frac{p_z}{E_n+m_e} I_{n-1}(\xi_-)
\\ -\frac{\sqrt{2neB}}{E_n+m_e} I_n(\xi_-)
\end{array}\right)\,, \qquad u^-(\vec{p}_{\nysc}, n, y) =  \left(
\begin{array}{c} 0 \\ I_{n}(\xi_-) \\
- \frac{\sqrt{2neB}}{E_n+m_e} I_{n-1}(\xi_-) \\
-\frac{p_z}{E_n+m_e} I_n(\xi_-)\end{array}\right)\,,} and
\eq{\label{eq:vspinor} v^+(\vec{p}_{\nysc}, n, y) =  \left(
\begin{array}{c} \frac{p_z}{E_n+m_e} I_{n-1}(\xi_+)
\\ \frac{\sqrt{2neB}}{E_n+m_e} I_n(\xi_+) \\
I_{n-1}(\xi_+) \\ 0 \end{array}\right)\,, \qquad
v^-(\vec{p}_{\nysc}, n, y) =  \left(
\begin{array}{c} \phantom{-} \frac{\sqrt{2neB}}{E_n+m_e} I_{n-1}(\xi_+) \\
- \frac{p_z}{E_n+m_e} I_{n}(\xi_+) \\
0 \\ I_n(\xi_+) \end{array}\right)\,.}
\end{subequations}
The $I_m(\xi)$ are functions of the Hermite polynomials %
\eq{\label{eq:Im} I_m(\xi) = \left(\frac{\sqrt{eB}}{2^m\, m!\,
\sqrt{\pi}}\right)^{1/2} e^{- \xi^2 / 2}\, H_m(\xi)} %
where the dimensionless parameter $\xi$ is defined by %
\eq{\label{eq:xi} \xi_\pm = \sqrt{eB}\,y \pm
\frac{p_x}{\sqrt{eB}}\,.}

Recall that the Hermite polynomials $H_m(\xi)$ are only defined for
nonnegative values of $m$. Therefore, we must define $I_{-1}(\xi) =
0 $.  This means that the electron in the lowest Landau energy level
($n=0$) cannot exist in spin-up state and the positron in the lowest
Landau energy level cannot exist in the spin-down state.

The normalization in Eq.~(\ref{eq:Im}) has been chosen such that the
functions $I_m(\xi)$ obey the following delta-function
representation \cite[p.~86]{Arfken:1995a}
\begin{eqnarray}
\delta(y - y') &=& \frac{\delta(\xi - \xi')}{\left|\partial
y / \partial{\xi}\right|} \nonumber\\
&=& \sqrt{eB}\, \delta(\xi - \xi') \nonumber \\
&=& \sqrt{eB}\, \sum_{n=0}^{\infty} \frac{1}{2^n\, n!\,
\sqrt{\pi}}\, e^{- \xi^2 / 2}\, H_n(\xi)\, e^{- \xi'^2 / 2}\,
H_n(\xi') \nonumber\\
\delta(y - y') &=& \sum_{n=0}^{\infty} I_n(\xi) I_n(\xi')\,.
\end{eqnarray}

For convenience we choose to normalize our 1-particle states in a
``box'' with dimensions $L_x L_y L_z = V$ such that the states are
defined as
\begin{subequations}\label{eq:state}
\begin{equation}\label{eq:estate}
|e\rangle = \left| \vec{p}_{1 \nysc}, n_1, s_1 \right\rangle =
\frac{1}{\sqrt{L_x Lz}}\, \hat{a}^{s_1\, \dag}_{e\,{\vec{p}}_{1
\nysc}, n_1}\, |0\rangle \,\phantom{,}
\end{equation}
\begin{equation}\label{eq:pstate}
|\bar{e}\rangle = \left| \vec{p}_{2 \nysc}, n_2, s_2 \right\rangle
= \frac{1}{\sqrt{L_x Lz}}\, \hat{b}^{s_2\, \dag}_{e\,{\vec{p}}_{2
\nysc}, n_2}\, |0\rangle \,,
\end{equation}
\end{subequations}
and the completeness relation for the states is
\begin{equation}\label{eq:comrel}
1 = \sum_{n=0}^{\infty} \sum_{s=\pm} \int\frac{\rmd^2
\vec{p}_{\nysc}}{(2\pi)^2} L_x L_z \left| \vec{p}_{\nysc}, n, s
\right\rangle \left\langle \vec{p}_{\nysc}, n, s \right|\,.
\end{equation}

\subsection{Spin sums\label{ssec:ss}}
In order to evaluate the production rate for our process, we must
derive the completeness relations for summations over the spin of
the fermions.  For a detailed calculation of the rules see
reference~\cite{Bhattacharya:2004nj}. The results of the calculation
are as follows
\begin{subequations}\label{eq:ss}
\begin{eqnarray}\label{eq:ssu}
\sum_{s=+,-} u^s(\vec{p}_{\nysc}, n, y') \bar{u}^s(\vec{p}_{\nysc},
n, y) &=& \left(2(E_n+m_e)\right)^{-1} \biggl\{\ \left[ m_e (1 -
\sigma^3) + \notp{p}_{\parallel} + \notp{q}_{\parallel} \gamma^5
\right]
I_{n}(\xi_-') I_n(\xi_-) \nonumber\\
&& \phantom{\left(2(E_n+m_e)\right)^{-1}} + \left[ m_e (1 +
\sigma^3) + \notp{p}_{\parallel} - \notp{q}_{\parallel} \gamma^5
\right] I_{n-1}(\xi_-') I_{n-1}(\xi_-) \nonumber\\
&& \phantom{\left(2(E_n+m_e)\right)^{-1}} + \sqrt{2neB}
\left(\gamma^1 + i \gamma^2\right) I_{n-1}(\xi_-') I_{n}(\xi_-) \nonumber\\
&& \phantom{\left(2(E_n+m_e)\right)^{-1}} + \sqrt{2neB}
\left(\gamma^1 - i \gamma^2\right) I_{n}(\xi_-') I_{n-1}(\xi_-)
\biggr\}
\end{eqnarray}
and
\begin{eqnarray}\label{eq:ssv}
\sum_{s=\pm} v^s(\vec{p}_{\nysc}, n, y) \bar{v}^s(\vec{p}_{\nysc},
n, y') &=& \left(2(E_n+m_e)\right)^{-1} \biggl\{\ \left[ - m_e (1 -
\sigma^3) + \notp{p}_{\parallel} + \notp{q}_{\parallel} \gamma^5
\right]
I_{n}(\xi_+) I_n(\xi_+') \nonumber\\
&& \phantom{\left(2(E_n+m_e)\right)^{-1}} + \left[ - m_e (1 +
\sigma^3) + \notp{p}_{\parallel} - \notp{q}_{\parallel} \gamma^5
\right] I_{n-1}(\xi_+) I_{n-1}(\xi_+') \nonumber\\
&& \phantom{\left(2(E_n+m_e)\right)^{-1}} - \sqrt{2neB}
\left(\gamma^1 + i \gamma^2\right) I_{n-1}(\xi_+) I_{n}(\xi_+') \nonumber\\
&& \phantom{\left(2(E_n+m_e)\right)^{-1}} - \sqrt{2neB}
\left(\gamma^1 - i \gamma^2\right) I_{n}(\xi_+) I_{n-1}(\xi_+')
\biggr\}\,,
\end{eqnarray}
\end{subequations}
where
\begin{eqnarray}\label{eq:pqp}
p_{\parallel}^\mu &=& (E, 0 , 0, p_z) \\
q_{\parallel}^\mu &=& (p_z, 0 , 0, E)\,.
\end{eqnarray}

The above results have been derived using the standard ``Bjorken and
Drell'' representation for the $\gamma$-matrices
\cite{Bjorken:1964a}
\begin{equation}\label{eq:gammas}
\gamma^0 = \left( \begin{array}{cc} 1 & 0 \\ 0 & -1 \end{array}
\right)\,, \qquad \gamma^i = \left( \begin{array}{cc} 0 & \sigma^i
\\ -\sigma^i & 0 \end{array}\right)\,.
\end{equation}

\subsection{Neutrino field operator}\label{ssec:nfo}
Having no charge, the neutrino's field operator solution
$\psi_\nu(x)$ is not modified due to the magnetic field.  We present
it here for easy reference
\begin{equation}\label{eq:nfieldop}
\psi_{\nu}(x) = \sum_{s=\pm} \int\frac{\rmd^3 \vec{p}}{(2\pi)^3}
\left[ u^s(p)\, e^{- i p \cdot x}\, \hat{a}^s_{\nu} + v^s(p)\, e^{+
i p \cdot x}\, {\hat{b}^{s\, \dag}_{\nu}} \right]\,,
\end{equation}
where the creation and annihilation operators obey the
conventional anticommutation relations
\begin{equation}\label{eq:nanticomm}
\left\{ \hat{a}^s_{\nu}, \hat{a}^{s'\, \dag}_{\nu}\right\} = \left\{
\hat{b}^s_{\nu}, \hat{b}^{s'\, \dag}_{\nu}\right\} = (2\pi)^3\,
\delta^{ss'}\, \delta^3(\vec{p} - \vec{p}\,')\,.
\end{equation}
The neutrino bi-spinors follow the standard spin sum rules
\begin{equation}\label{eq:nspinsum}
\sum_{s = \pm} u^s(p) \bar{u}^s(p) = \sum_{s = \pm} v^s(p)
\bar{v}^s(p) = \notp{p}\,,
\end{equation}
where we take the Standard Model neutrino mass to be zero.

With ``box'' normalization the 1-particle states for the neutrino
are
\begin{equation}\label{eq:nstate}
|\nu \rangle = \left| \vec{p}, s \right\rangle =
\frac{1}{\sqrt{V}}\, \hat{a}^{s\, \dag}_{\nu\,{\vec{p}}}\,
|0\rangle\, ,
\end{equation}
satisfying the completeness relation
\begin{equation}\label{eq:ncomrel}
1 = \sum_{s=\pm} \int\frac{\rmd^3 \vec{p}}{(2\pi)^3} V \left|
\vec{p}, s \right\rangle \left\langle \vec{p}, s \right|\,.
\end{equation}

\section{The production rate\label{sec:prodrate}}
The quantity of interest for the process $\nu \to \nu e \bar{e}$ in
a background magnetic field is the rate at which the
electron-positron pairs are produced $\Gamma$. The production rate
is defined as the probability per unit time for creation of pairs
\begin{equation}\label{eq:prate}
\Gamma = \lim_{T \to \infty} \frac{\mathcal{P}}{T}\,.
\end{equation}
where $T$ is the timescale on which the process is normalized. We
begin by finding the probability $\mathcal{P}$ of our reaction
\begin{equation}\label{eq:prob}
\mathcal{P} = \sum_{n_1, n_2 = 0}^{\infty} \int \frac{\rmd^3
\vec{p}'}{(2\pi)^3}\, V\, \int \frac{\rmd^2 \vec{p}_{1
\nysc}}{(2\pi)^2}\, L_x L_z\, \int \frac{\rmd^2 \vec{p}_{2
\nysc}}{(2\pi)^2}\, L_x L_z\, \sum_{s,s',s_1,s_2} \left|
\left\langle \vec{p}', s'; \vec{p}_{1 \nysc}, n_1, s_1; \vec{p}_{2
\nysc}, n_2, s_2 \left| \hat{S} \right| \vec{p}, s \right\rangle
\right|^2\,.
\end{equation}
In Eq.~\ref{eq:prob} quantities with the index $1$ correspond to the
electron, those with index $2$ to the positron, the primed
quantities to the final neutrino, and the unprimed quantities
correspond to the initial neutrino.

\subsection{The scattering matrix}\label{ssec:scatmat}
The scattering matrix
\begin{equation}\label{eq:Smat}
S = \left\langle \vec{p}', s'; \vec{p}_{1 \nysc}, n_1, s_1;
\vec{p}_{2 \nysc}, n_2, s_2 \left| \hat{S} \right| \vec{p}, s
\right\rangle
\end{equation}
naturally depends on the flavor of the neutrino.  While the
process involving the electron neutrino can advance through either
the charged ($W$) or neutral ($Z$) current, the muon (or tau)
neutrino can only proceed through the latter.  For this reason we
will break the scattering matrix into a neutral component
\begin{subequations}\label{eq:Smatzw}
\begin{equation}\label{eq:Smatz}
S_Z = \left\langle \vec{p}', s'; \vec{p}_{1 \nysc}, n_1, s_1;
\vec{p}_{2 \nysc}, n_2, s_2 \left| \hat{S}_Z \right| \vec{p}, s
\right\rangle
\end{equation}
and a charged component
\begin{equation}\label{eq:Smatw}
S_W = \left\langle \vec{p}', s'; \vec{p}_{1 \nysc}, n_1, s_1;
\vec{p}_{2 \nysc}, n_2, s_2 \left| \hat{S}_W \right| \vec{p}, s
\right\rangle \,,
\end{equation}
\end{subequations}
where the scattering operators are defined by the Standard Model
Lagrangian as
\begin{subequations}\label{eq:SopZW}
\begin{equation}\label{eq:SopZ}
\hat{S}_Z = \frac{e^2}{2^3\, \cos^2\theta_W\, \sin^2\theta_W} \int
\rmd^4x\, \overline{\psi}_e(x) \gamma^{\mu} \left(g_V^e - g_A^e
\gamma^5\right) \psi_e(x) Z_\mu(x) \int \rmd^4x'\,
\overline{\psi}_{\nu_l}(x') \gamma^{\sigma} \left(1 -
\gamma^5\right) \psi_{\nu_l}(x') Z_{\sigma}(x')
\end{equation}
\begin{equation}\label{eq:SopW}
\hat{S}_W = \frac{e^2}{2^3\, \sin^2\theta_W} \int \rmd^4x\,
\overline{\psi}_e(x) \gamma^{\mu} \left(1 - \gamma^5\right)
\psi_{\nu_e}(x) W^-_\mu(x) \int \rmd^4x'\,
\overline{\psi}_{\nu_e}(x') \gamma^{\sigma} \left(1 -
\gamma^5\right) \psi_{e}(x') W^+_{\sigma}(x')\,,
\end{equation}
\end{subequations}
and $\theta_W$ is the weak-mixing angle, $\nu_l$ indicates a
neutrino of any flavor, $\nu_e$ refers to a electron neutrino, and
the vector and axial vector couplings for the electron are
\begin{subequations}\label{eq:coups}
\begin{eqnarray}
g_V^e &=& -\frac{1}{2} + 2 \sin^2\theta_W \label{eq:coupsv}\\
g_A^e &=& -\frac{1}{2}\,.\label{eq:coupsa}
\end{eqnarray}
\end{subequations}
In our analysis we will be using incoming neutrino energies that
are well below the rest energies of the $Z$ and $W$ bosons.
Therefore, we can safely make the 4-fermion effective coupling
approximation to the $Z$ and $W$ propagators
\begin{subequations}\label{eq:4fermprop}
\begin{eqnarray}\label{eq:4fermpropz}
\left\langle0\left| T\left( Z_{\mu}(x)
Z_{\sigma}(x')\right)\right|0\right\rangle & \to & \delta^4(x -
x')\, \frac{g_{\mu\sigma}}{m_Z^2} \\
\left\langle0\left| T\left( W^-_{\mu}(x) W^+_{\sigma}(x')
\right)\right|0\right\rangle & \to & \delta^4(x - x')\,
\frac{g_{\mu\sigma}}{m_W^2}\,.\label{eq:4fermpropw}
\end{eqnarray}
\end{subequations}
After making this approximation our expressions for the scattering
operators simplify to
\begin{subequations}\label{eq:SopZWb}
\begin{eqnarray}
\hat{S}_Z &=& \frac{G_F}{\sqrt{2}} \int \rmd^4x\,
\overline{\psi}_e(x) \gamma^{\mu} \left(g_V^e - g_A^e
\gamma^5\right) \psi_e(x)\, \overline{\psi}_{\nu_l}(x)
\gamma_{\mu} \left(1 - \gamma^5\right)
\psi_{\nu_l}(x) \qquad \label{eq:SopZb}\\
\hat{S}_W &=& \frac{G_F}{\sqrt{2}} \int \rmd^4x\,
\overline{\psi}_e(x) \gamma^{\mu} \left(1 - \gamma^5\right)
\psi_{\nu_e}(x)\, \overline{\psi}_{\nu_e}(x) \gamma_{\mu} \left(1 -
\gamma^5\right) \psi_{e}(x)\,,\label{eq:SopWb}
\end{eqnarray}
\end{subequations}
where $G_F/\sqrt{2} = e^2 / (8\, \sin^2\theta_W\, m_W^2)$, and we
have made use of the fact that $\cos^2\theta_W = m_W^2 / m_Z^2$.

After substituting of the scattering operators
(Eqs.~(\ref{eq:SopZWb})) into the expressions for the components of
the scattering matrix (Eqs.~(\ref{eq:Smatzw})), we can use our
results from sections~\ref{ssec:efo} and \ref{ssec:nfo} to write the
components in the form of
\begin{equation}\label{eq:scatmatform}
S_{Z/W} = \frac{i (2\pi)^3\, \delta^3\left(p_{\nysc} - p_{\nysc}' -
p_{\nysc,1} - p_{\nysc,2}\right)}{L_x\, L_z\, V}\,
\mathcal{M}_{Z/W}\,,
\end{equation}
where
\begin{subequations}
\begin{eqnarray}
\mathcal{M}_Z &=& \frac{-i G_F}{2^2\sqrt{2}}\, \sqrt{\frac{ (E_{n_1}
+ m_e) (E_{n_2} + m_e)}{E E' E_{n_1} E_{n_2}}}\, \bar{u}^{s'}(p')
\gamma_\mu \left(1 - \gamma^5\right) u^{s}(p)
\nonumber\\
&& \times \int \rmd y\, e^{i (p_y - p'_y)y}\,
\bar{u}^{s_1}\left(\vec{p}_{1 \nysc}, n_1, y\right) \gamma^\mu
\left(g_V^e - g_A^e \gamma^5\right) v^{s_2}\left(\vec{p}_{2
\nysc}, n_2, y\right) \qquad \label{eq:Mz}\\
\mathcal{M}_W &=& \frac{i G_F}{2^2\sqrt{2}}\, \sqrt{\frac{ (E_{n_1}
+ m_e) (E_{n_2} + m_e)}{E E' E_{n_1} E_{n_2}}}\, \bar{u}^{s'}(p')
\gamma_\mu \left(1 - \gamma^5\right) v^{s_2}\left(\vec{p}_{2 \nysc},
n_2, y\right)
\nonumber\\
&& \times \int \rmd y\, e^{i (p_y - p'_y)y}\,
\bar{u}^{s_1}\left(\vec{p}_{1 \nysc}, n_1, y\right) \gamma^\mu
\left(1 - \gamma^5\right) u^{s}(p) \,. \label{eq:Mw}
\end{eqnarray}
\end{subequations}
The reversal of sign on Eq.~(\ref{eq:Mw}) relative to
Eq.~(\ref{eq:Mz}) is from the anticommutation of the field
operators.  The scattering amplitude for the charged component
$\mathcal{M}_W$ can be transformed into the form of the neutral
component $\mathcal{M}_Z$ by making use of a Fierz rearrangement
formula
\begin{equation}\label{eq:Fierz}
\bar{u}_1 \gamma_\mu \left(1 - \gamma^5\right) u_2\, \bar{u}_3
\gamma^\mu \left(1 - \gamma^5\right) u_4 = - \bar{u}_1 \gamma_\mu
\left(1 - \gamma^5\right) u_4\, \bar{u}_3 \gamma^\mu \left(1 -
\gamma^5\right) u_2\,,
\end{equation}
such that
\begin{eqnarray}
\mathcal{M}_W &=& \frac{-i G_F}{2^2\sqrt{2}}\, \sqrt{\frac{ (E_{n_1}
+ m_e) (E_{n_2} + m_e)}{E E' E_{n_1} E_{n_2}}}\, \bar{u}^{s'}(p')
\gamma_\mu \left(1 - \gamma^5\right) u^{s}(p)
\nonumber\\
&& \times \int \rmd y\, e^{i (p_y - p'_y)y}\,
\bar{u}^{s_1}\left(\vec{p}_{1 \nysc}, n_1, y\right) \gamma^\mu
\left(1 - \gamma^5\right) v^{s_2}\left(\vec{p}_{2 \nysc}, n_2,
y\right) \,.\qquad \label{eq:Mwb}
\end{eqnarray}

With the rearrangement of $\mathcal{M}_W$ in Eq.~(\ref{eq:Mwb}), we
can now express the scattering amplitude in terms of the type of
incoming neutrino. The muon neutrino can only proceed through
exchange of a $Z$-boson, so its scattering amplitude is just that of
$\mathcal{M}_Z$
\begin{eqnarray}\label{eq:Mnumu}
\mathcal{M}_{\nu_\mu} &=& \mathcal{M}_Z \nonumber\\
\mathcal{M}_{\nu_\mu} &=& \frac{i G_F}{2^3\sqrt{2}}\, \sqrt{\frac{
(E_{n_1} + m_e) (E_{n_2} + m_e)}{E E' E_{n_1} E_{n_2}}}\,
\bar{u}^{s'}(p') \gamma_\mu \left(1 - \gamma^5\right) u^{s}(p)
\nonumber\\
&& \times \int \rmd y\, e^{i (p_y - p'_y)y}\,
\bar{u}^{s_1}\left(\vec{p}_{1 \nysc}, n_1, y\right) \gamma^\mu
\left(G_V^{-} - \gamma^5\right) v^{s_2}\left(\vec{p}_{2 \nysc}, n_2,
y\right)\,. \qquad
\end{eqnarray}
The scattering matrix for a tau neutrino, and the subsequent decay
rate, is exactly the same as the muon neutrino.  We will keep the
notation as $\nu_\mu$ for simplicity.

The electron neutrino has both a $Z$-boson exchange component and
an $W$-boson exchange component.  Therefore we must add the
amplitudes to find its scattering amplitude
\begin{eqnarray}\label{eq:Mnue}
\mathcal{M}_{\nu_e} &=& \mathcal{M}_Z + \mathcal{M}_W\nonumber\\
\mathcal{M}_{\nu_e} &=& \frac{-i G_F}{2^3\sqrt{2}}\, \sqrt{\frac{
(E_{n_1} + m_e) (E_{n_2} + m_e)}{E E' E_{n_1} E_{n_2}}}\,
\bar{u}^{s'}(p') \gamma_\mu \left(1 - \gamma^5\right) u^{s}(p)
\nonumber\\
&& \times \int \rmd y\, e^{i (p_y - p'_y)y}\,
\bar{u}^{s_1}\left(\vec{p}_{1 \nysc}, n_1, y\right) \gamma^\mu
\left(G_V^{+} - \gamma^5\right) v^{s_2}\left(\vec{p}_{2 \nysc}, n_2,
y\right)\,.
\end{eqnarray}
Note that the scattering amplitudes for electron (Eq.~\ref{eq:Mnue})
and non-electron neutrinos (Eq.~\ref{eq:Mnumu}) depend on a
generalized vector coupling $G_V$ defined by
\begin{equation}\label{eq:GVpm}
G_V^{\pm} = 1 \pm 4 \sin^2\theta_W\,.
\end{equation}

We see that the scattering amplitudes for an incoming electron
neutrino versus an incoming muon neutrino differ only in the value
of the generalized vector coupling and an overall sign.  And the
overall sign will be rendered meaningless once the amplitude is
squared.  Therefore, we choose to make no distinction between the
two processes, other than keeping the generalized vector coupling as
$G_V^{\pm}$, until we discuss the results in
section~\ref{sec:results}.

\subsection{The form of the production rate\label{ssec:Gform}}
Having determined the scattering matrix $S$ and scattering amplitude
$\mathcal{M}$ in section~\ref{ssec:scatmat}, we can now make series
of substitutions of those results to find the expression for the
production rate $\Gamma$.  We begin by substituting the form of the
scattering matrix (Eq.~(\ref{eq:scatmatform})) into the expression
for the production rate (Eq.~(\ref{eq:prate})
\begin{eqnarray}\label{eq:prateb}
\Gamma &=& \lim_{T \to \infty} \frac{\mathcal{P}}{T} \nonumber\\
&=& \lim_{T,V \to \infty} T^{-1}\, \sum_{n_1, n_2 = 0}^{\infty} \int
\frac{\rmd^3 \vec{p}'}{(2\pi)^3}\, V\, \int \frac{\rmd^2 \vec{p}_{1
\nysc}}{(2\pi)^2}\, L_x L_z\, \int \frac{\rmd^2 \vec{p}_{2
\nysc}}{(2\pi)^2}\, L_x L_z \sum_{s,s',s_1,s_2} \left| \left\langle
\vec{p}', s'; \vec{p}_{1 \nysc}, n_1, s_1; \vec{p}_{2 \nysc}, n_2,
s_2 \left| \hat{S} \right| \vec{p}, s \right\rangle
\right|^2 \nonumber\\
&=& \lim_{T,V \to \infty} T^{-1}\, \sum_{n_1, n_2 = 0}^{\infty} \int
\frac{\rmd^3 \vec{p}'}{(2\pi)^3}\, V\, \int \frac{\rmd^2 \vec{p}_{1
\nysc}}{(2\pi)^2}\, L_x L_z\, \int \frac{\rmd^2 \vec{p}_{2
\nysc}}{(2\pi)^2}\, L_x L_z \sum_{s,s',s_1,s_2} \left| \frac{i
(2\pi)^3\, \delta^3\left(p_{\nysc} - p_{\nysc}' - p_{\nysc,1} -
p_{\nysc,2}\right)}{L_x\, L_z\, V}\, \mathcal{M}\right|^2 \nonumber\\
\Gamma &=& \lim_{T,V \to \infty} (2\pi TV)^{-1}\, \sum_{n_1, n_2 =
0}^{\infty} \int \rmd^3 \vec{p}'\, \int \rmd^2 \vec{p}_{1 \nysc}\,
\int \rmd^2 \vec{p}_{2 \nysc} \left(\delta^3(p_{\nysc} - p_{\nysc}'
- p_{\nysc,1} - p_{\nysc,2})\right)^2 \overline{\left| \mathcal{M}
\right|^2}\,,\qquad
\end{eqnarray}
where $\overline{\left| \mathcal{M} \right|^2}$ is the square of the
scattering amplitude after summing over spins
\begin{equation}\label{eq:M2avg}
\overline{\left| \mathcal{M} \right|^2} = \sum_{s,s',s_1,s_2} \left|
\mathcal{M}\right|^2\,.
\end{equation}

We can simplify the square of the 3-dimensional delta function by
expressing one of the 3-dimensional delta functions as a series of
integrals over space-time coordinates
\begin{equation}\label{eq:delta2}
\left(\delta^3(p_{\nysc} - p_{\nysc}' - p_{\nysc,1} -
p_{\nysc,2})\right)^2 = \delta^3(p_{\nysc} - p_{\nysc}' -
p_{\nysc,1} - p_{\nysc,2}) \, \int \frac{\rmd^3 \nysc}{(2\pi)^3}\,
e^{i (p - p' - p_1 - p_2) \cdot\, \nysc}\,.
\end{equation}
By using the remaining set of delta functions to reduce the
exponential to unity, we can write the integrand in terms of the
dimensions of our normalization ``box''
\begin{eqnarray}\label{eq:delta2b}
\left(\delta^3(p_{\nysc} - p_{\nysc}' - p_{\nysc,1} -
p_{\nysc,2})\right)^2 &=& \delta^3(p_{\nysc} - p_{\nysc}' -
p_{\nysc,1} - p_{\nysc,2}) \, \int \frac{\rmd^3
\nysc}{(2\pi)^3} \nonumber\\
\left(\delta^3(p_{\nysc} - p_{\nysc}' - p_{\nysc,1} -
p_{\nysc,2})\right)^2 &=& \delta^3(p_{\nysc} - p_{\nysc}' -
p_{\nysc,1} - p_{\nysc,2}) \, \frac{T L_x L_z}{(2\pi)^3}\,.
\end{eqnarray}
With the above result for the square of the delta function, the
production rate in Eq.~(\ref{eq:prateb}) simplifies to
\begin{equation}\label{eq:drate}
\Gamma = \lim_{L_y \to \infty} \sum_{n_1, n_2 = 0}^{\infty} \int
\rmd^3 \vec{p}'\, \int \rmd^2 \vec{p}_{1 \nysc}\, \int \rmd^2
\vec{p}_{2 \nysc}\, \delta^3(p_{\nysc} - p_{\nysc}' - p_{\nysc,1} -
p_{\nysc,2}) \frac{\overline{\left| \mathcal{M} \right|^2}}{(2\pi)^4
L_y}\,.
\end{equation}

The square of the scattering amplitude goes as the product of two
traces
\begin{eqnarray}\label{eq:M2avgb}
\overline{\left| \mathcal{M} \right|^2} &=& \sum_{s,s',s_1,s_2} \left| \mathcal{M}\right|^2 \nonumber\\
&=& \frac{G_F^2}{2^7}\, \frac{ (E_{n_1}
+ m_e) (E_{n_2} + m_e)}{E E' E_{n_1} E_{n_2}} \nonumber\\
&& \times \sum_{s,s'} \bar{u}^{s}(p) \gamma_\sigma \left(1 -
\gamma^5\right) u^{s'}(p') \bar{u}^{s'}(p') \gamma_\mu \left(1 -
\gamma^5\right) u^{s}(p) 
\int \rmd y\, e^{i (p_y - p'_y)y}\, \int \rmd y'\,
e^{-i (p_y - p'_y)y'}\nonumber\\
&& \times \sum_{s_1, s_2} \bar{v}^{s_2}\left(\vec{p}_{2 \nysc}, n_2,
y'\right) \gamma^\sigma \left(G_V^{\pm} - \gamma^5\right)
u^{s_1}\left(\vec{p}_{1 \nysc}, n_1, y'\right)
 \bar{u}^{s_1}\left(\vec{p}_{1 \nysc}, n_1, y\right) \gamma^\mu
\left(G_V^{\pm} - \gamma^5\right) v^{s_2}\left(\vec{p}_{2 \nysc},
n_2, y\right)\nonumber
\end{eqnarray}
\begin{eqnarray}\label{eq:M2avgc}
\overline{\left| \mathcal{M} \right|^2} &=& \frac{G_F^2}{2^{9}}\,
\left(E E' E_{n_1} E_{n_2}\right)^{-1}\, \int \rmd y\, e^{i (p_y -
p'_y)y}\, \int \rmd y'\, e^{-i (p_y - p'_y)y'} \mathrm{Tr} \left\{
\gamma_\sigma \left(1 - \gamma^5\right) \notp{p}' \gamma_\mu \left(1
-
\gamma^5\right) \notp{p} \right\} \nonumber\\
&& \times \mathrm{Tr}\biggl\{ %
\gamma^\sigma \left(G_V^{\pm} - \gamma^5\right)%
\Bigl[ \left( m_e (1 - \sigma^3) + %
\notp{p}_{1 \parallel} + \notp{q}_{1 \parallel} \gamma^5 \right)%
I_{n_1}(\xi_{-,1}') I_{n_1}(\xi_{-,1}) \nonumber\\
&& \phantom{\times \mathrm{Tr}\  %
\gamma^\sigma \left(G_V^{\pm} - \gamma^5\right)}%
+ \sqrt{2 n_1 eB} \left(\gamma^1 + i \gamma^2\right) %
I_{n_1 - 1}(\xi_{-,1}') I_{n_1}(\xi_{-,1}) \nonumber\\
&& \phantom{\times \mathrm{Tr}\  %
\gamma^\sigma \left(G_V^{\pm} - \gamma^5\right)}%
+ \sqrt{2 n_1 eB} \left(\gamma^1 - i \gamma^2\right) %
I_{n_1}(\xi_{-,1}') I_{n_1 - 1}(\xi_{-,1}) \nonumber\\
&& \phantom{\times \mathrm{Tr}\  %
\gamma^\sigma \left(G_V^{\pm} - \gamma^5\right)}%
+ \left( m_e (1 + \sigma^3) + %
\notp{p}_{1 \parallel} - \notp{q}_{1 \parallel} \gamma^5 \right)%
I_{n_1 - 1}(\xi_{-,1}') I_{n_1 -1}(\xi_{-,1}) \Bigr] \nonumber\\
&& \phantom{\mathrm{Tr}\biggl\{} \times %
\gamma^\mu \left(G_V^{\pm} - \gamma^5\right)%
\Bigl[ \left( - m_e (1 - \sigma^3) + %
\notp{p}_{2 \parallel} + \notp{q}_{2 \parallel} \gamma^5 \right)%
I_{n_2}(\xi_{+,2}') I_{n_2}(\xi_{+,2}) \nonumber\\
&& \phantom{\times \mathrm{Tr}\  %
\gamma^\sigma \left(G_V^{\pm} - \gamma^5\right)}%
- \sqrt{2 n_2 eB} \left(\gamma^1 + i \gamma^2\right) %
I_{n_2 - 1}(\xi_{+,2}') I_{n_2}(\xi_{+,2}) \nonumber\\
&& \phantom{\times \mathrm{Tr}\  %
\gamma^\sigma \left(G_V^{\pm} - \gamma^5\right)}%
- \sqrt{2 n_2 eB} \left(\gamma^1 - i \gamma^2\right) %
I_{n_2}(\xi_{+,2}') I_{n_2 - 1}(\xi_{+,2}) \nonumber\\
&& \phantom{\times \mathrm{Tr}\  %
\gamma^\sigma \left(G_V^{\pm} - \gamma^5\right)}%
+ \left(- m_e (1 + \sigma^3) + %
\notp{p}_{2 \parallel} - \notp{q}_{2 \parallel} \gamma^5 \right)%
I_{n_2 - 1}(\xi_{+,2}') I_{n_2 -1}(\xi_{+,2}) \Bigr]\biggr\}, \nonumber\\
&&
\end{eqnarray}
where we have used our result for the summations over spin from
Eqs.~(\ref{eq:ss}) and (\ref{eq:nspinsum}).

The space-time dependence of Eq.~(\ref{eq:M2avgc}) can be factored
into terms like
\begin{equation}\label{eq:Inm}
I_{n, m} = \int \rmd y\, e^{i (p_y - p'_y)y}\, I_{n}(\xi_{-,1})\,
I_m(\xi_{+,2})
\end{equation}
and
\begin{equation}\label{eq:Inm*}
I^\ast_{n, m} = \int \rmd y'\, e^{-i (p_y - p'_y)y'}\,
I_{n}(\xi'_{-,1})\, I_m(\xi_{+,2}')\,,
\end{equation}
where the $I_{n,m}$ are functions of the momenta in the problem.

We have included a detailed calculation for the general form of
$I_{n,m}$ in appendix~\ref{ap:Inm}, but we only present the result
here
\begin{equation}\label{eq:Inma}
I_{n,m} = \left\{ \begin{array}{l l} %
\sqrt{\frac{n!}{m!}}\, e^{-\eta^2/2}\, e^{i \phi_0} \left(\eta_x + i
\eta_y\right)^{m-n} L_n^{m-n}\left(\eta^2\right)\,, & m
\geq n \geq 0 \\
\sqrt{\frac{m!}{n!}}\, e^{-\eta^2/2}\, e^{i \phi_0} \left(-\eta_x +
i \eta_y\right)^{n-m} L_m^{n-m}\left(\eta^2\right)\,, & n \geq m
\geq 0
\end{array}\right\}\,
\end{equation}
where
\begin{eqnarray}\label{eq:vars}
\eta_x &=& \frac{p_{1 x} + p_{2 x}}{\sqrt{2eB}} \\
\eta_y &=& \frac{p_{y} - p_{y}'}{\sqrt{2eB}} \\
\phi_0 &=& \frac{(p_y - p_y')(p_1 - p_2)}{2eB} \\
\eta^2 &=& \eta_x^2 + \eta_y^2\,,
\end{eqnarray}
and $L_n^{m-n}(\eta^2)$ are the associated Laguerre polynomials.

The full results of the traces and their subsequent contraction are
nontrivial but have been included in appendix~\ref{ap:trace}. It is
important to note, however, that the only dependence on the
$x$-components of the electron and positron momentum is that which
appears in Eq.~(\ref{eq:Inma}) for $I_{n,m}$. Furthermore, we notice
that all terms in the averaged square of the scattering amplitude
have factors that go as a product of $I_{n,m}$ and $I^\ast_{n',m'}$.
Therefore, the coefficient $e^{i\, \phi_0}$ in Eq.~(\ref{eq:Inma})
will vanish when this product is taken. The only remaining
$x$-dependence of these two momenta appear as their sum in the
parameter $\eta_x = (p_{1 x} + p_{2 x})/\sqrt{2eB}$. This helps to
simplify the phase-space integral for our production rate
(Eq.~(\ref{eq:drate})) which is proportional to
\begin{equation}\label{eq:gampspropa}
\Gamma \propto \lim_{L_y \to \infty} \frac{1}{L_y} \int \rmd
p_{1\,x}\, \int \rmd p_{2\,x} \,.
\end{equation}
If we make a change of variable from the $x$-component of the
positron momentum $p_{2\,x}$ to the parameter $\eta_x$, the
relationship in Eq.~(\ref{eq:gampspropa}) is rewritten as
\begin{equation}\label{eq:gampspropb}
\Gamma \propto \lim_{L_y \to \infty} \frac{\sqrt{2eB}}{L_y} \int
\rmd p_{1\,x}\, \int \rmd \eta_x \,.
\end{equation}
Because there is no longer any explicit dependence on the
$x$-component of the electron's momentum $p_{1,x}$ in the averaged
square of our scattering amplitude, we can simply evaluate the
integral
$$
\int \rmd p_{1,x}\,.
$$
To evaluate this integral we must determine its limits.  As
discussed previously, we have elected to use ``box'' normalization
on our states.  This means that our particle is confined to a large
box with dimensions $L_x$, $L_y$, and $L_z$. The careful reader will
note that we have already taken the limit that these dimensions go

to infinity in some places, particularly in Eq.~(\ref{apInmeq:4}),
but it is imperative that we be cautious here, as we could naively
evaluate the integral over $p_{1,x}$ to be infinite.

Physically, the charged particles in our final state act as harmonic
oscillators circling about the magnetic field lines. While they are
free to slide about the lines along the $z$-axis, the particles are
confined to circular orbits in the $x$ and $y$-directions no larger
than the dimensions of the box.  For a charged particle undergoing
circular motion in a constant magnetic field, the $x$-component of
momentum is related to the $y$-position vector by
\begin{equation}
p_x = - eQ B y\,
\end{equation}
where $Q$ is the charge of the particle in units of the proton
charge $e=|e|$.  Therefore, the limits on $p_{1,x}$ are proportional
to the limits on the size of our box in the $y$-direction.  The
integral over the electron's momentum in the $x$-direction is
\begin{equation}
\int_{-eBL_y/2}^{eBL_y/2} \rmd p_{1,x} = eBL_y\,,
\end{equation}
and the result helps to cancel the factor of $L_y$ that already
appears in the form of the production rate.  We can now safely take
the limit that our box has infinite size, and the production rate
now has the form
\begin{equation}\label{eq:drateb}
\Gamma = \sum_{n_1, n_2 = 0}^{\infty} \int \rmd^3 \vec{p}'\, \int
\rmd \vec{p}_{1 z}\, \int \rmd \vec{p}_{2 z}\, \int \rmd \eta_x\,
\sqrt{2eB}\, \delta^3(p_{\nysc} - p_{\nysc}' - p_{\nysc,1} -
p_{\nysc,2})\, \frac{eB\, \overline{\left| \mathcal{M}
\right|^2}}{(2\pi)^4}\,.
\end{equation}

\section{Results}\label{sec:results}
In our expression for the total production rate
(Eq.~(\ref{eq:drateb})), one will notice is that there is a sum over
all possible values of the Landau levels. As a consequence of energy
conservation, upper limits do exist for the summation over the
electron's Landau level $n_1$
\begin{eqnarray}\label{eq:constr1}
E &=& E' + E_{n_1} + E_{n_2} \nonumber\\
E &\geq& E_{n_1} + m_e \nonumber\\
E - m_e &\geq& \sqrt{m_e^2 + 2 n_1 eB} \nonumber\\
n_1 &\leq& \frac{E (E - 2 m_e)}{2 e B}\,,
\end{eqnarray}
and a similar one for the positron's Landau level
\begin{eqnarray}\label{eq:constr2}
E &=& E' + E_{n_1} + E_{n_2} \nonumber\\
E &\geq& \sqrt{m_e^2 + 2 n_1 e B} + E_{n_2} \nonumber\\
E - \sqrt{m_e^2 + 2 n_1 e B} &\geq& \sqrt{m_e^2 + 2 n_2 eB} \nonumber\\
n_2 &\leq& \frac{\left(E - \sqrt{m_e^2 + 2 n_1 e B}\right)^2 -
m_e^2}{2 e B}\,.
\end{eqnarray}
These relationships help to constrain the extent of the summations.
Physically, these constraints can be thought of as limits on the
size of the electron's (or positron's) effective mass, where the
electron (or positron) occupying the $n^{\mathrm{th}}$ Landau level
has an effective mass
\begin{equation}
m_\ast = \sqrt{m_e^2 + 2neB}\,
\end{equation}
and energy
\begin{equation}
E_n = \sqrt{p_z^2 + {m_\ast}^2 }\,.
\end{equation}
For low incoming neutrino energies and large magnetic field
strengths ($eB > m_e^2$), the constraints put very tight bounds on
the limits of the summations.  However, higher incoming energies and
low magnetic field strengths impose limits that still require a
great deal of computation time.  For instance, at threshold ($E = 2
m_e$) there can exist only one possible configuration of Landau
levels ($n_1 = n_2 = 0$), while at an energy ten times that of
threshold and a magnetic field equal to the critical field ($B =
B_\mathrm{c} = m_e^2/e = 4.414 \times 10^{13}~\mathrm{G}$) there are
nearly 7000 possible states. At the same magnetic field but an
energy that is 100 times that of threshold, there are almost 70
million states. However, for incoming neutrino energies less than a
certain value
\begin{equation}\label{eq:E00lim}
E < m_e + \sqrt{m_e^2 + 2 e B}
\end{equation}
only the lowest Landau level is occupied, $n_1\,, n_2 = 0$.  And
even at energies above, yet near, this value we expect that
production of electrons and positrons in the $n_1\,, n_2 = 0$ level
is still the dominant mode of production because it has more phase
space available.

\begin{figure}[h]
\subfigure[Incoming electron neutrino
\label{fig:G00vE-e}]{\includegraphics{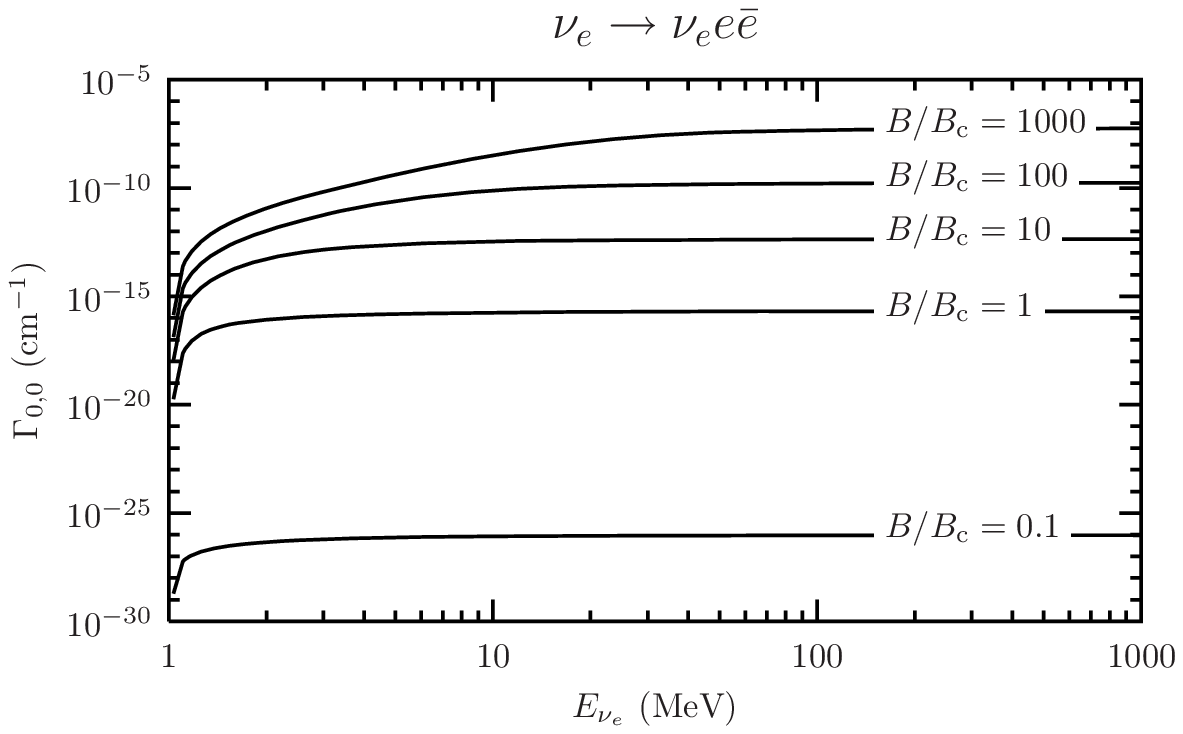}} \vfill
\subfigure[Incoming muon (tau)
neutrino\label{fig:G00vE-mu}]{\includegraphics{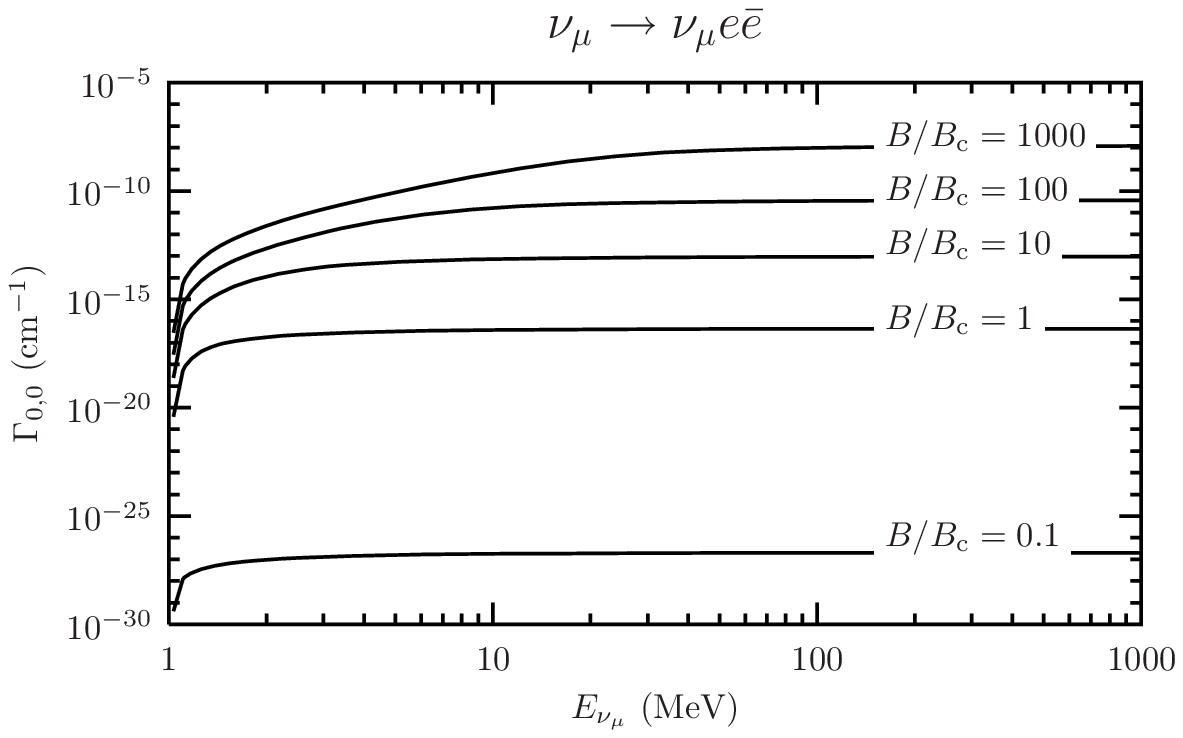}}
\caption[Production rates for the $n_1\,, n_2 = 0 $ Landau level as
a function of incoming neutrino energy.]{Production rates for the
$n_1\,, n_2 = 0 $ Landau levels where $\Gamma$ is the rate of
production, $E_\nu$ is the energy of the incoming neutrino, and the
magnetic field is measured relative to the critical field
$B_\mathrm{c} = 4.414 \times 10^{13}~\mathrm{G}$. All plots are for
a neutrino that is perpendicularly incident to the magnetic
field.\label{fig:G00vE}}
\end{figure}

Production rates at the $0,0$ Landau level are presented in
FIG.~\ref{fig:G00vE} for both the electron and muon neutrinos.  (All
of the results for muon-type neutrinos are valid for tau-type
neutrinos.) One interesting feature of these results is the
flattening out of the rates at higher energies. The energy region at
which this flattening begins increases with increasing magnetic
field strength, and it appears to be in the neighborhood of energies
just above the limit set in Eq.~(\ref{eq:E00lim}).  At energies in
this regime we expect that modes of production into other Landau
levels are stimulated, which helps to explain why the behavior of
the $0,0$ production rates change above this area.

\begin{figure}[h]
\includegraphics{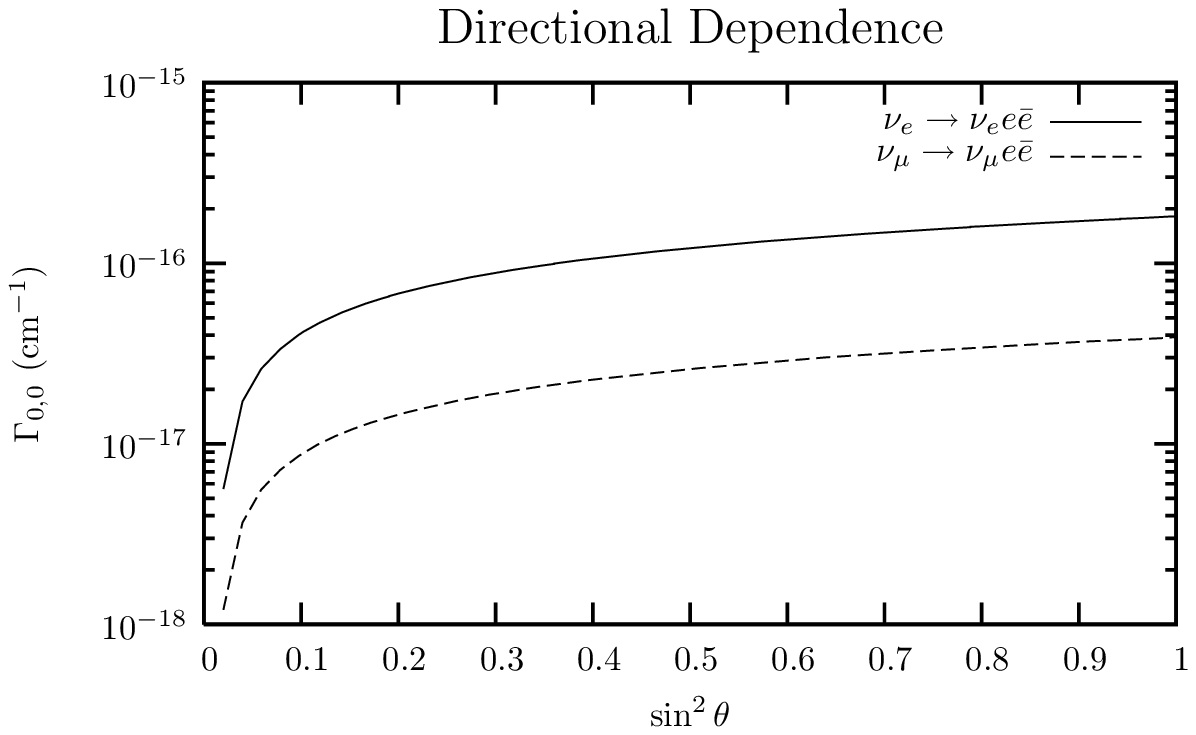} \caption[Directional
dependence of the incoming neutrino on the production rate.]{The
production rate's dependance on the direction of the incoming
neutrino..  The production rate is for the $0,0$ Landau level with
an electron of energy $E_\nu = 20 m_e$ traveling at an angle
$\theta$ relative to a magnetic field of strength equal to the
critical field $B= B_\mathrm{c}$. Data is included for both an
incoming electron-type neutrino (solid line) and a muon-type
neutrino (dashed line). If we average over $\theta$, then the
average production rate is $1.38\times 10^{-16}~\mathrm{cm}^{-1}$
for electron-type neutrinos or $2.94 \times
10^{-17}~\mathrm{cm}^{-1}$ for muon or tau-type.\label{fig:zdep}}
\end{figure}

We should note that the results given in this work are all for an
incoming neutrino traveling transversely to the magnetic field. The
rates are maximized in this case as can be seen in the example found
in FIG.~\ref{fig:zdep} for an initial electron neutrino with energy
$E_{\nu_e} = 20 m_e$ in a magnetic field equal to the critical field
$B = B_\mathrm{c} = m_e^2/e$.

\begin{figure}[h]
\centering \subfigure[Incoming electron neutrino
\label{fig:G01vE-e}]{\includegraphics{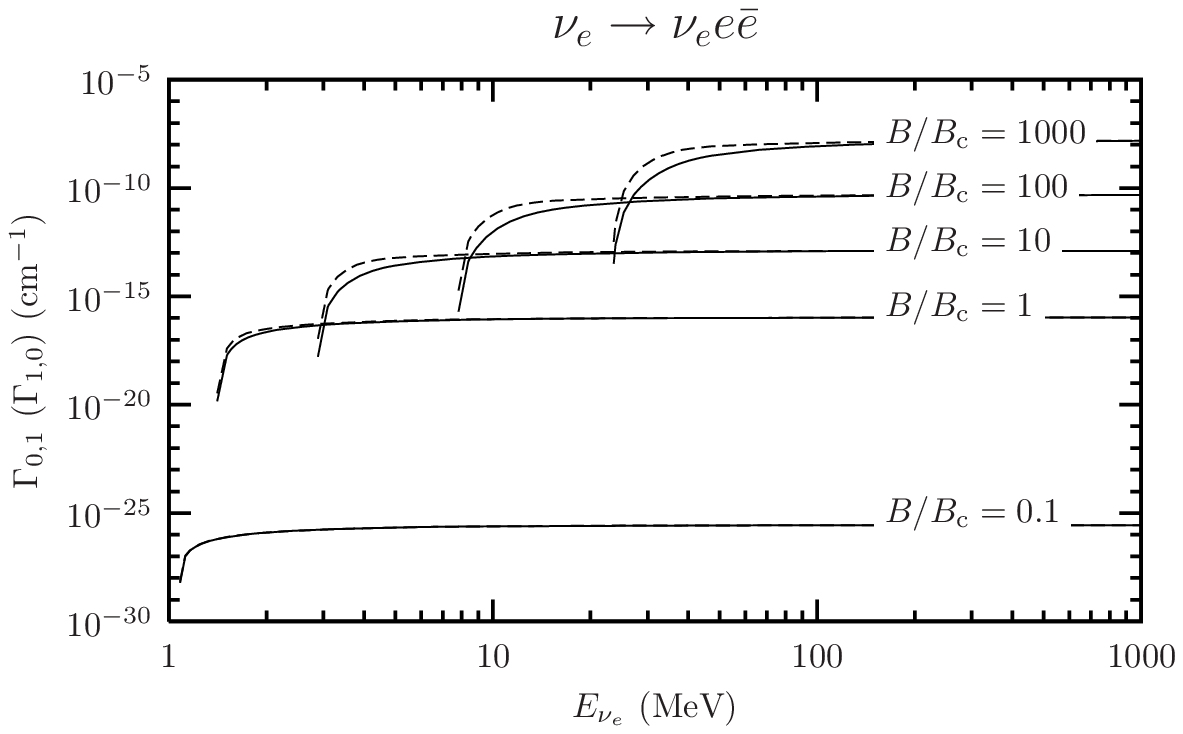}} \vfill
\subfigure[Incoming muon
neutrino\label{fig:G01vE-mu}]{\includegraphics{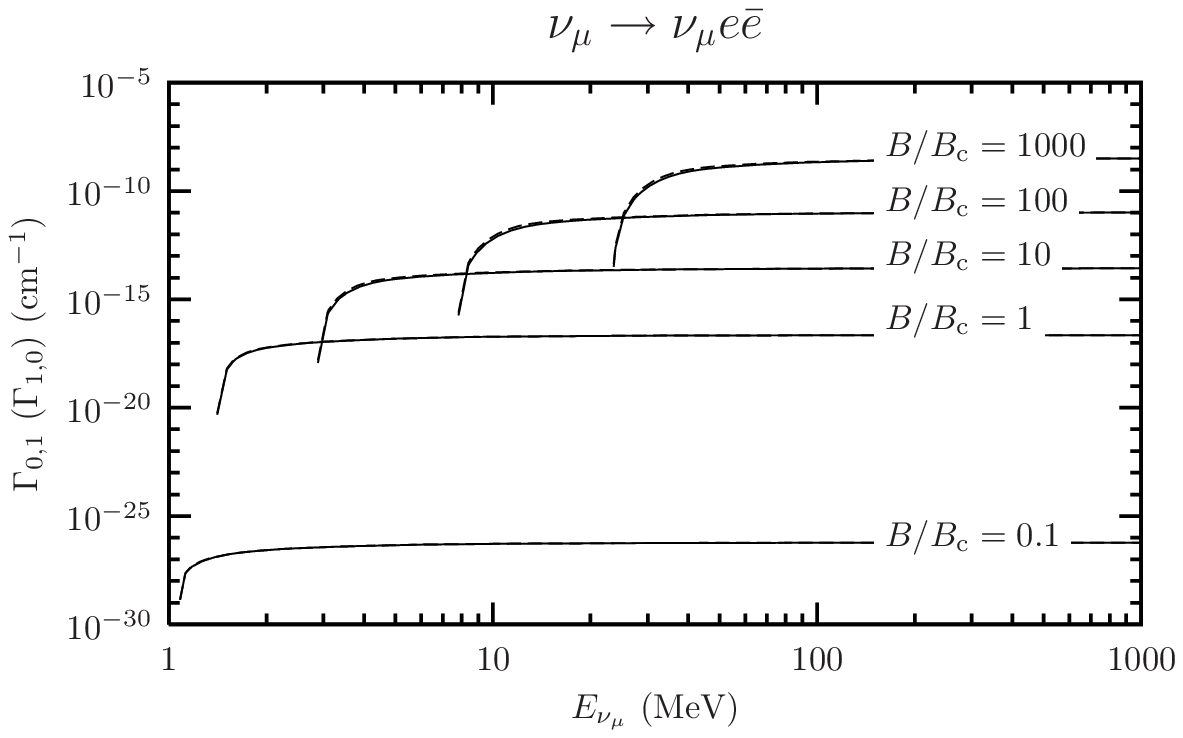}}
\caption[Production rates for the $n_1 =0\,, n_2 = 1$ and $n_1 =1\,,
n_2 = 0$ Landau levels as function of energy.]{Production rates for
the $n_1 =0\,, n_2 = 1$ (solid) and $n_1 =1\,, n_2 = 0$ (dashed)
Landau levels where $\Gamma$ is the rate of production, $E_\nu$ is
the energy of the incoming neutrino, and the magnetic field is
measured relative to the critical field $B_\mathrm{c} = 4.414 \times
10^{13}~\mathrm{G}$.\label{fig:G01vE}}
\end{figure}

\begin{figure}[h]
\centering \subfigure[Incoming electron neutrino
\label{fig:G020vE-e}]{\includegraphics{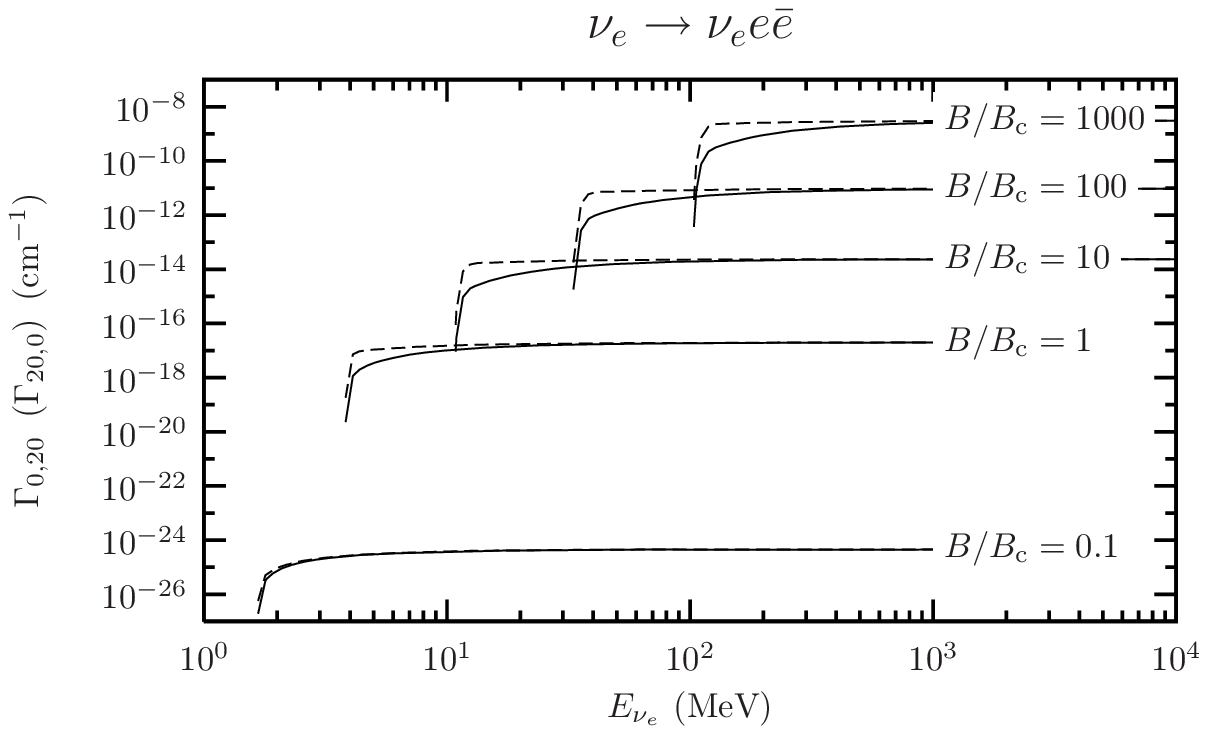}} \vfill
\subfigure[Incoming muon
neutrino\label{fig:G020vE-mu}]{\includegraphics{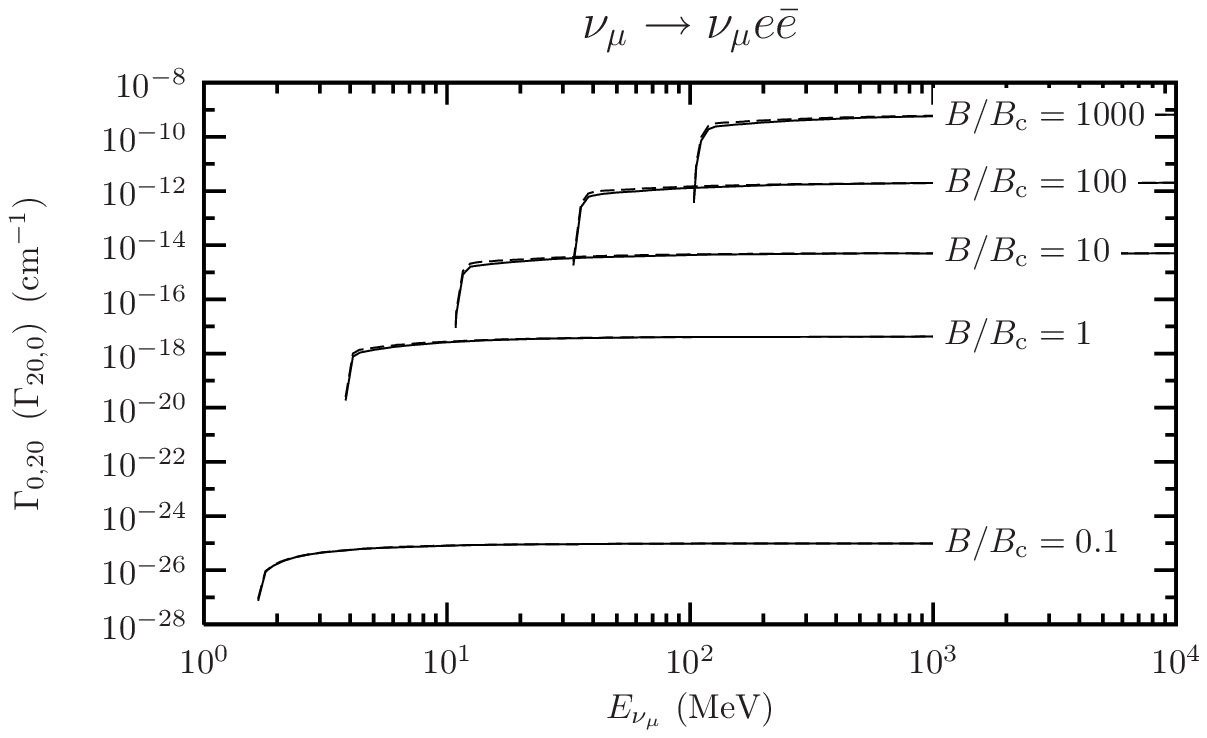}}
\caption[Production rates for the $n_1 =0\,, n_2 = 20$ and $n_1
=20\,, n_2 = 0$ Landau levels as a function of energy.]{Production
rates for the $n_1 =0\,, n_2 = 20$ (solid) and $n_1 =20\,, n_2 = 0$
(dashed) Landau levels where $\Gamma$ is the rate of production,
$E_\nu$ is the energy of the incoming neutrino, and the magnetic
field is measured relative to the critical field $B_\mathrm{c} =
4.414 \times 10^{13}~\mathrm{G}$.\label{fig:G020vE}}
\end{figure}

\begin{figure}[h]
\centering \subfigure[Incoming electron neutrino
\label{fig:G1010vE-e}]{\includegraphics{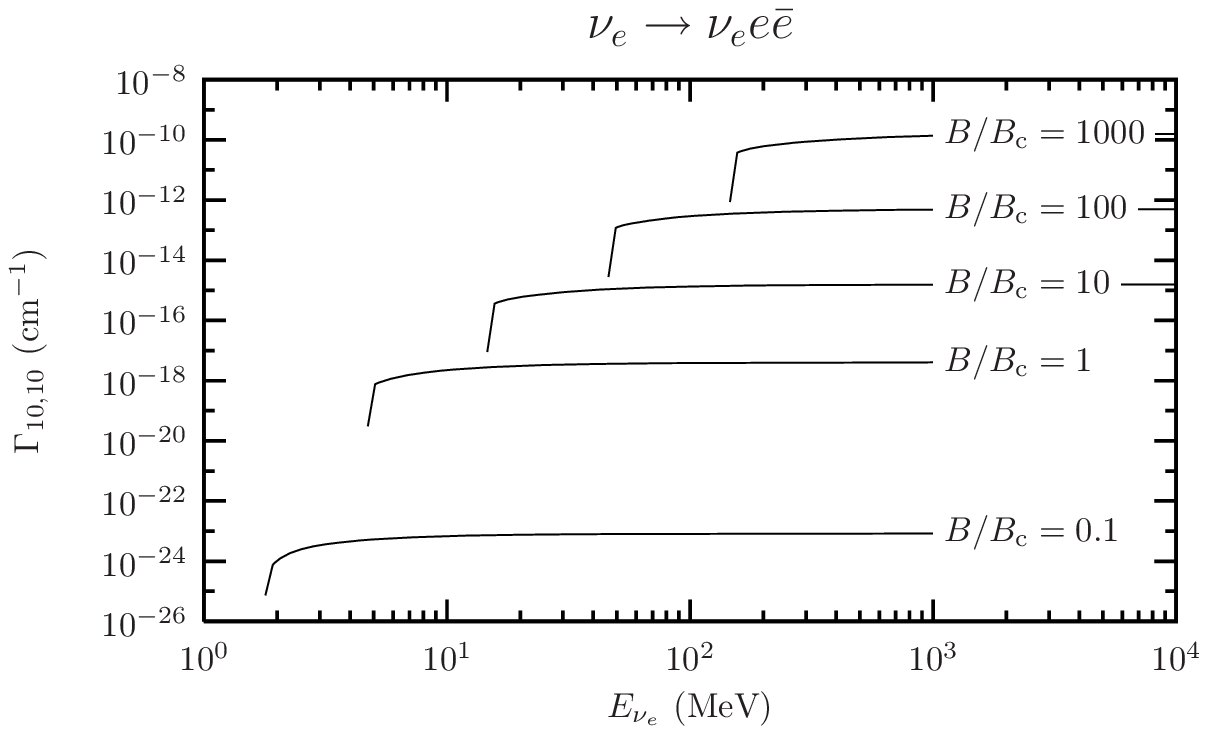}} \vfill
\subfigure[Incoming muon
neutrino\label{fig:G1010vE-mu}]{\includegraphics{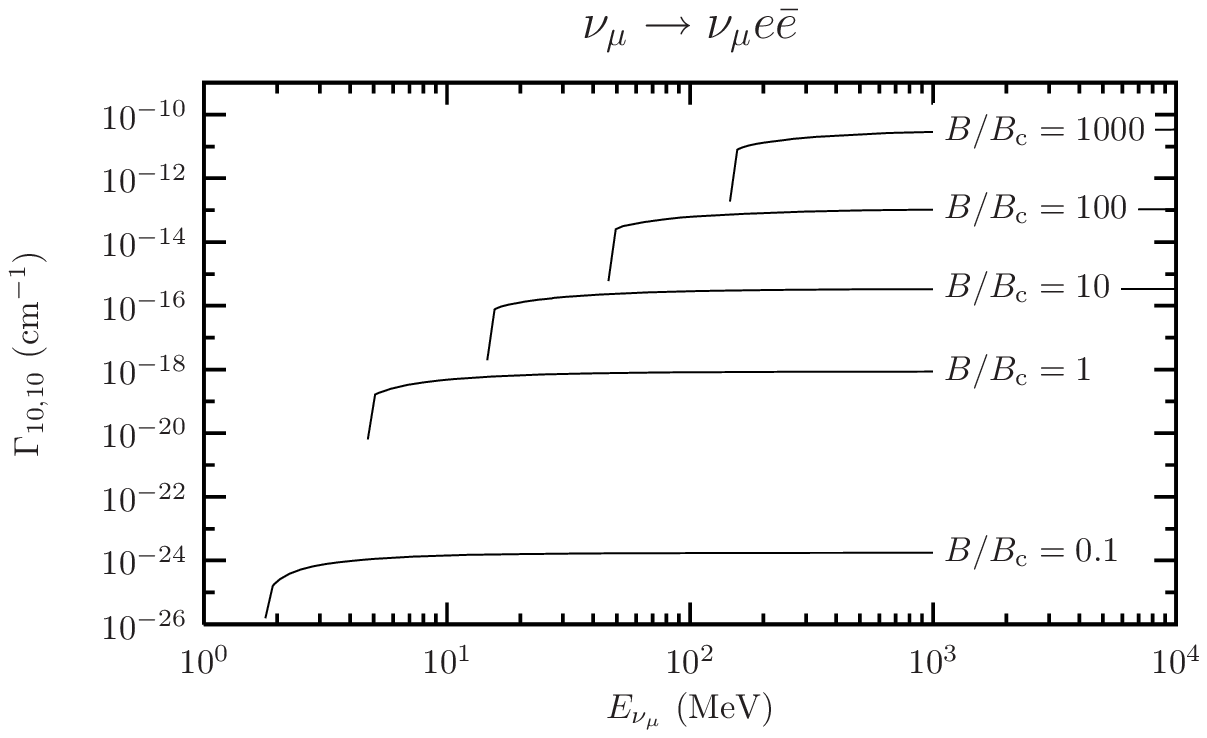}}
\caption[Production rates for the $n_1\,, n_2 = 10 $ Landau level as
a function of energy.]{Production rates for the $n_1\,, n_2 = 10 $
Landau levels where $\Gamma$ is the rate of production, $E_\nu$ is
the energy of the incoming neutrino, and the magnetic field is
measured relative to the critical field $B_\mathrm{c} = 4.414 \times
10^{13}~\mathrm{G}$.\label{fig:G1010vE}}
\end{figure}

For comparison purposes, the production rates for other combinations
of Landau levels have been calculated.  These include the $1,0$ and
$0,1$ cases (FIG.~\ref{fig:G01vE}), the $20,0$ and $0,20$ cases
(FIG.~\ref{fig:G020vE}), and the $10,10$ case
(FIG.~\ref{fig:G1010vE}).  The first noteworthy feature of these
results is that the production rates are decreasing at higher Landau
levels. Because the energy required to create the pair goes as
\begin{eqnarray*}
E_{\mathrm{pair}} &=& E_{n_1} + E_{n_2} \\
&=& \sqrt{p_{1\,z}^2 + 2 n_1 e B + m_e^2} + \sqrt{p_{2\,z}^2 + 2 n_2
e B + m_e^2}
\\
E_{\mathrm{pair}} &\geq& \sqrt{2 n_1 e B + m_e^2} + \sqrt{2 n_2 e B
+ m_e^2}\,,
\end{eqnarray*}
the available phase space for the process should decrease in the
order $0,0\ \to\ 0,1\ \to\ 0,20\ \to\ 10,10$.  And as can be seen in
FIGS.~\ref{fig:G00vE}, \ref{fig:G01vE}, \ref{fig:G020vE}, and
\ref{fig:G1010vE}, the production rates fall off accordingly.

Another interesting feature of these results is the apparent
preference for the creation of electrons in the highest of the two
Landau levels.  That is, the rate of production is larger for the
state $n_1 = i\,, n_2 = 0$ than for $n_1 = 0\,, n_2 = i$
(FIGS.~\ref{fig:G01vE} and \ref{fig:G020vE}). This behavior is
especially significant over the range of incoming neutrino energies
near its threshold value for creating pairs in the given states.
Though the $i,0$ production rate is larger and increases more
quickly in this ``near-threshold'' range than its $0,i$ counterpart,
both curves plateau at higher energies, and their difference
approaches zero.  This difference is presumably caused by the
positron having to share the $W$'s energy with the final
electron-type neutrino. This also explains why such an effect is not
seen for muon and tau-type neutrinos that only proceed through the
neutral current reaction.

\begin{figure}[h]
\centering \subfigure[Incoming electron neutrino
\label{fig:Gapprox-e}]{\includegraphics{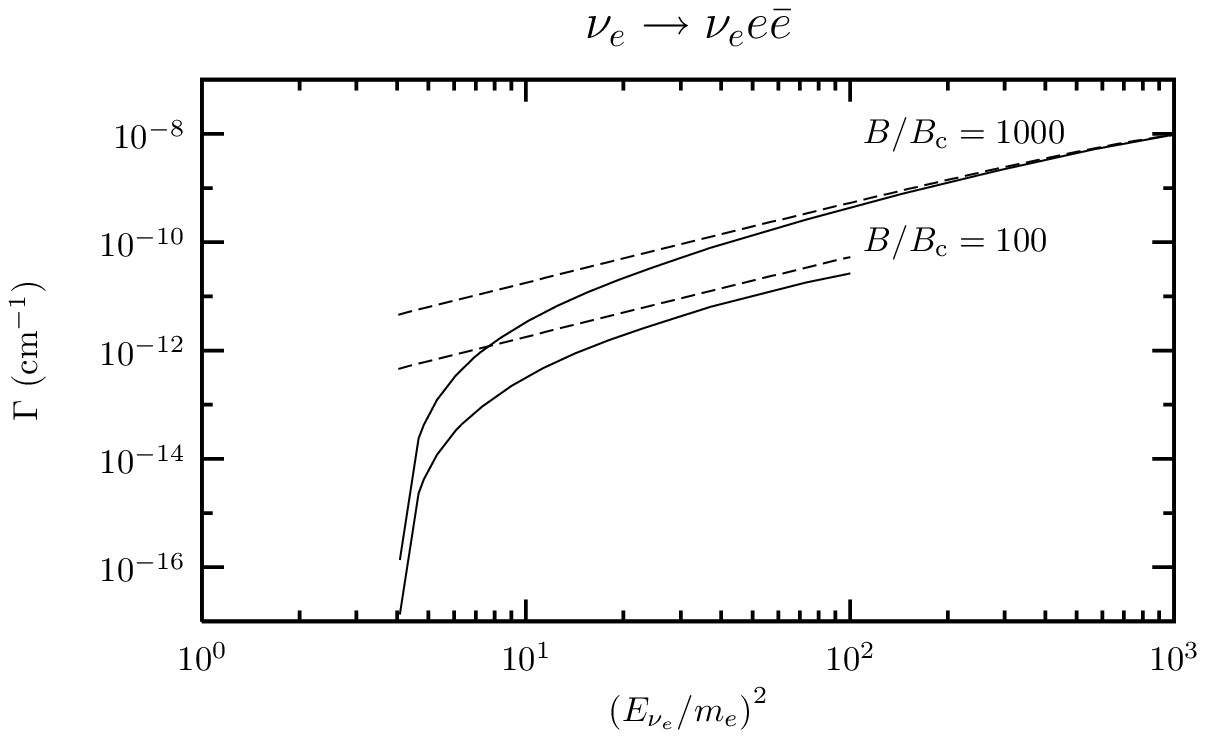}} \vfill
\subfigure[Incoming muon
neutrino\label{fig:Gapprox-mu}]{\includegraphics{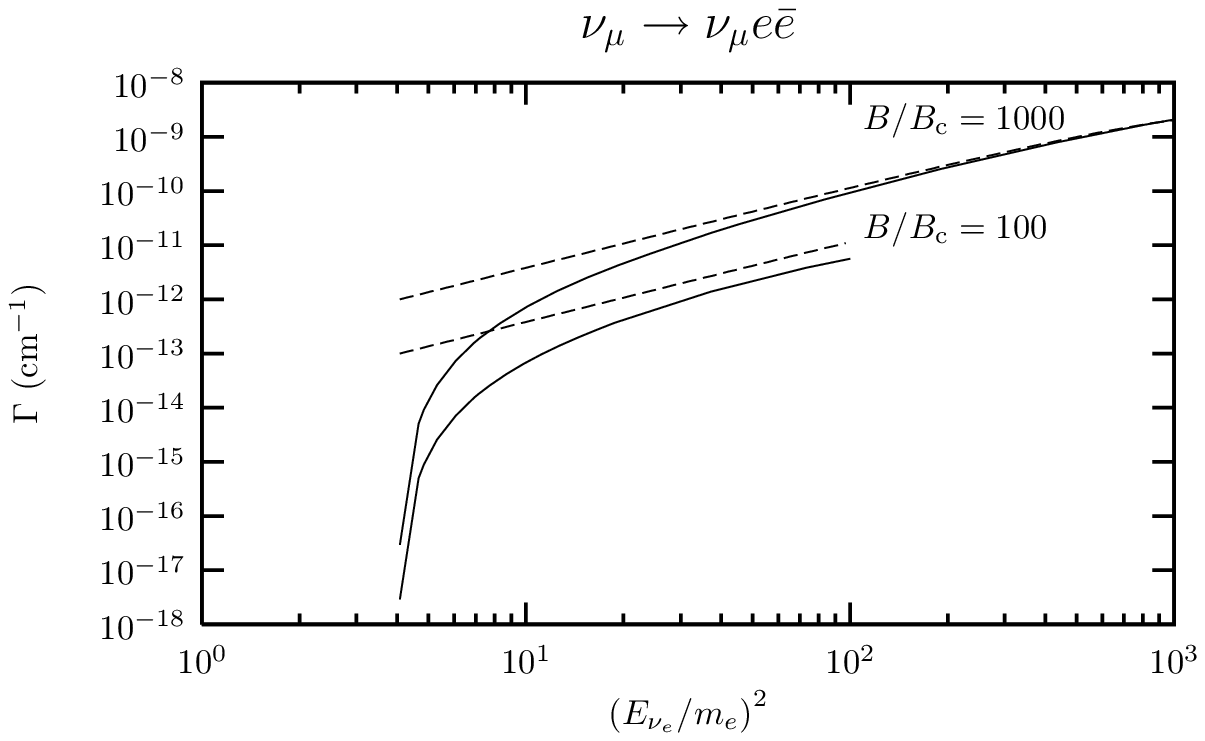}}
\caption[The total production rate and its approximation.]{The total
production rate (solid lines) and its approximation (dashed
lines)\cite{Kuznetsov:1996vy} for energies and magnetic field
satisfying the condition $E_\nu^2 < eB$. $E_\nu$ is the energy of
the incoming neutrino, $m_e$ is the mass of the electron, and the
magnetic field is measured relative to the critical field
$B_\mathrm{c} = 4.414 \times
10^{13}~\mathrm{G}$.\label{fig:Gapprox}}
\end{figure}

It was mentioned in section~\ref{sec:intro} that previous authors
have considered this process under two limiting
cases~\cite{Kuznetsov:1996vy}. One is when the square of the energy
of the initial-state neutrino and the magnetic field strength
satisfy the conditions $E_\nu \gg eB \gg m_e^2$. Under these
conditions many possible Landau levels could be stimulated, offering
a multitude of production modes. Therefore, it would be
inappropriate to compare their expression to our results for a
specific set of Landau levels.  However, the second limiting case is
for $eB > E_\nu^2 \gg m_e^2$.  This condition is slightly more
restrictive than our condition for the energies below which only the
lowest energy Landau levels are occupied (Eq.~(\ref{eq:E00lim})). In
this regime our results for the $0,0$ state are the total production
rates, and we can compare our results to the expression derived by
the previous authors~\cite{Kuznetsov:1996vy}
\begin{equation}\label{eq:Kuzaprox}
\Gamma = \frac{G_F^2 \left({G_V^\pm}^2 + 1\right)}{2^6 \pi^3}\, eB\,
E_\nu^3\, \left( 1 + \mathcal{O}\left(E^2/eB\right)\right)\,,
\end{equation}
where we have taken the direction of the incoming neutrino to be
perpendicular to the magnetic field's direction. Results of this
comparison are shown in FIG.~\ref{fig:Gapprox}.

The results in FIG.~\ref{fig:Gapprox} demonstrate the drawbacks of
using the approximation in Eq.~(\ref{eq:Kuzaprox}). While the
expression is very simple, it gives only reasonable agreement with
the production rate at a magnetic field equal to 100 times that of
the critical field ($B = 100\, B_\mathrm{c}$). Here it
overestimates, at the very least, by a factor of two, and the
inclusion of higher order corrections makes no significant
improvement. One reason for the disagreement at this field strength
is that there is only a very small range of energies that satisfy
the condition $eB > E_\nu^2 \gg m_e^2$. Therefore at higher field
strengths we should get better agreement, and we do.  Closer
inspection of FIG.~\ref{fig:Gapprox} reveals that the differences
are less than a factor of three for neutrino energies in the range
$2~\mathrm{MeV} < E_\nu < 20~\mathrm{MeV}$, and the expression
successfully provides a good order of magnitude estimation.  Though
the estimate will improve at higher magnetic field strengths, it
begins to loose relevance as there are only a handful of known
objects (namely magnetars) that can conceivably possess fields as
high as $10^{15}$~G. Even for these objects, fields stronger than
$10^{15}$~G cause instability in the star and the field begins to
diminish~\cite{Thompson:1993a}.

Probing the limiting case $E_\nu \gg \sqrt{eB}$ is imperative
because our present work has already demonstrated nontrivial
deviation from approximate methods for realistic astrophysical
magnetic field strengths and neutrino energies near and below the
value $\sqrt{eB}$. But, as was mentioned previously, the number of
Landau level states which contribute to the total production rate
grows very rapidly in this higher energy regime, and we need to sum
over these states. Future work will attempt to do these sums by
using an approximation routine that can interpolate between rates
for known sets of Landau levels. This will provide a flexible way to
balance accuracy with computation time while determining when the
production rate deviates from its limiting behavior. The
significance of these deviations will only be known when a more
complete understanding of the role that neutrino processes play in
events such as supernova core-collapse and in the formation of the
resulting neutron star. This work aims to improve that
understanding.

\begin{acknowledgments}
It is our pleasure to thank Craig Wheeler for several discussions
about supernovae and Palash Pal for helping us to understand
Ref.~\cite{Bhattacharya:2002aj}.  This work was supported in part by
the U.S.~Department of Energy under Grant No.~DE-F603-93ER40757 and
by the National Science Foundation under Grant PHY-0244789 and
PHY-0555544.
\end{acknowledgments}

\bibliography{bib}

\appendix\pdfbookmark[0]{Appendices}{Ap}
\section{Calculation of $\mathbf{I_{n,m}}$\label{ap:Inm}}
In section~\ref{ssec:Gform} we discuss the fact that the squared
scattering amplitude has coefficients that are integrals over the
space-time coordinate $y$
\begin{equation}\label{apInmeq:1}
I_{n, m} = \int \rmd y\, e^{i (p_y - p'_y)y}\, I_{n}(\xi_{-,1})\,
I_m(\xi_{+,2})\,.
\end{equation}
In this appendix we will derive the result after integrating over
$y$.

By defining new parameters
\begin{eqnarray}\label{apInmeq:3}
\zeta &=& \sqrt{eB}\, y \\
\zeta_i &=& p_{i x} / \sqrt{eB} \\
\zeta_0 &=& (p_y - p'_y)/\sqrt{eB}
\end{eqnarray}
and using the definition of $\xi$ (Eq.~(\ref{eq:xi})) we can make a
change of variable from $y$ to $\zeta$ and rewrite $I_{n,m}$ as
\begin{equation}\label{apInmeq:4}
I_{n, m} = \int_{-\infty}^{\infty} \frac{\rmd \zeta}{\sqrt{eB}}\,
e^{i \zeta_0 \zeta}\, I_{n}(\zeta - \zeta_1)\, I_m(\zeta + \zeta_2)
\end{equation}
where the limits of integration are $\pm\infty$ because we have
taken the limit of $L_y$ as it approaches $\infty$. The $I_{n}(\xi)$
in Eq.~(\ref{eq:Im}) depend on the Hermite polynomials $H_n(\xi)$,
which can be represented as a contour integral in the following way
\cite[Eq.~(13.8)]{Arfken:1995a}
\begin{equation}\label{apInmeq:5}
H_n(\xi) = \frac{n!}{2\pi i} \oint \rmd t\, t^{-n-1}\, e^{-t^2 +
2tx}\,.
\end{equation}
Substituting this definition of the Hermite polynomial into
Eq.~(\ref{eq:Im}) allows us to write the $I_{n,m}$ as
\begin{eqnarray}\label{apInmeq:6}
I_{n, m} &=& \int_{-\infty}^{\infty} \frac{\rmd \zeta}{\sqrt{eB}}\,
e^{i \zeta_0\, \zeta}\, \left(\frac{\sqrt{eB}}{2^n\, n!\,
\sqrt{\pi}}\right)^{1/2} e^{- (\zeta - \zeta_1)^2 / 2}\, H_n(\zeta -
\zeta_1) \left(\frac{\sqrt{eB}}{2^m\, m!\, \sqrt{\pi}}\right)^{1/2}
e^{- (\zeta + \zeta_2)^2 / 2}\, H_m(\zeta
+ \zeta_2) \nonumber\\
&=& \left(2^{n+m}\, n!\, m!\, \pi\right)^{-1/2}\,
\int_{-\infty}^{\infty} \rmd \zeta\, e^{i \zeta_0\, \zeta}\, e^{-
(\zeta - \zeta_1)^2
/ 2}\, H_n(\zeta - \zeta_1) 
e^{- (\zeta + \zeta_2)^2 / 2}\, H_m(\zeta + \zeta_2)
\nonumber\\
&=& \left(2^{n+m}\, n!\, m!\, \pi\right)^{-1/2}\,
\int_{-\infty}^{\infty} \rmd \zeta\, e^{i \zeta_0\, \zeta}\, e^{-
(\zeta - \zeta_1)^2
/ 2}\, e^{- (\zeta + \zeta_2)^2 / 2} \nonumber\\
&& \times  \, \frac{n!}{2\pi i} \oint \rmd t\, t^{-n-1}\,
e^{-t^2 + 2t(\zeta - \zeta_1)} 
\frac{m!}{2\pi i} \oint \rmd s\, s^{-m-1}\, e^{-s^2 + 2s(\zeta +
\zeta_2)}\,.
\end{eqnarray}

Next, we isolate all of the $\zeta$ dependence, interchange the
order of the integrals, and perform the integration over $\zeta$
\begin{eqnarray}\label{apInmeq:7}
\mathrm{Int}_1 &=& \int_{-\infty}^{\infty} \rmd \zeta\, \exp\left(
-\zeta^2 + \zeta (\zeta_1 - \zeta_2 + i \zeta_0 + 2t +
2s)\right) \nonumber \\
&=& \sqrt{\pi}\ \exp\left(\left(\zeta_1 - \zeta_2 + i \zeta_0 + 2t +
2s\right)^2/4\right)\,.
\end{eqnarray}
Substitution of this result back into Eq.~(\ref{apInmeq:6}) gives
\begin{eqnarray}\label{apInmeq:8}
I_{n, m} &=& \left(2^{n+m}\, n!\, m!\right)^{-1/2}\,
e^{-\left((\zeta_1 + \zeta_2)^2 + \zeta_0^2\right)/4}\, e^{i
\zeta_0 (\zeta_1 - \zeta_2)/2} \nonumber\\
&& \times  \, \frac{n!}{2\pi i} \oint \rmd t\, t^{-n-1}\,
e^{t(-\zeta_1 - \zeta_2 + i \zeta_0)} \, \frac{m!}{2\pi i} \oint
\rmd s\, s^{-m-1}\, e^{s(\zeta_1 +
\zeta_2 + i \zeta_0)}\, e^{2st}\,. \nonumber\\
&&
\end{eqnarray}

If $m \geq n$, then we can perform the integration over $s$ first
\begin{eqnarray}\label{apInmeq:9}
\mathrm{Int}_2 &=& \frac{m!}{2\pi i} \oint \rmd s\, s^{-m-1}\,
e^{s(\zeta_1 + \zeta_2 + i \zeta_0)}\, e^{2st}\,. \nonumber\\
&=& \left. \frac{\rmd^m}{\rmd s^m}\,  e^{s(\zeta_1 + \zeta_2 +
i \zeta_0 + 2t)}\right|_{s=0}\, \nonumber\\
&=& (\zeta_1 + \zeta_2 + i \zeta_0 + 2t)^m\,,
\end{eqnarray}
such that
\begin{eqnarray}\label{apInmeq:10}
I_{n, m} &=& \left(2^{n+m}\, n!\, m!\right)^{-1/2}\,
\exp\left(-\left((\zeta_1 + \zeta_2)^2 + \zeta_0^2\right)/4\right)\,
\exp\left(i
\zeta_0 (\zeta_1 - \zeta_2)/2\right) \nonumber\\
&& \times  \, \frac{n!}{2\pi i} \oint \rmd t\, \frac{(\zeta_1 +
\zeta_2 + i \zeta_0 + 2t)^m}{t^{n+1}}\, \exp\left(t(-\zeta_1 -
\zeta_2 + i \zeta_0)\right) \,.
\end{eqnarray}

The integration over $t$ is made easier by making the following
changes of variable
\begin{eqnarray}\label{apInmeq:11}
\eta_x &=& \frac{\zeta_1 + \zeta_2}{\sqrt{2}}\ =\ \frac{p_{1 x} +
p_{2 x}}{\sqrt{2eB}} \\
\eta_y &=& \frac{\zeta_0}{\sqrt{2}}\ =\ \frac{p_y -
p_y'}{\sqrt{2eB}} \\
\phi_0 &=& \frac{\zeta_0\, (\zeta_1 - \zeta_2)}{2} \ =\ \frac{(p_y
- p_y')(p_1 - p_2)}{2eB}\\
\eta^{\pm} &=& \eta_x \pm i \eta_y \\
\eta^2 &=& \eta^+\, \eta^-\ =\ \eta_x^2 + \eta_y^2 \\
t &=& \left(u - \eta^+\right)/\sqrt{2}\,.
\end{eqnarray}
The integration over the variable $t$ can now be written as
\begin{eqnarray}\label{apInmeq:12}
\mathrm{Int}_3 &=& \frac{n!}{2\pi i} \oint \frac{\rmd u}{\sqrt{2}}\,
\frac{(\sqrt{2}\, u)^m}{\left(\left(u -
\eta^+\right)/\sqrt{2}\right)^{n+1}}\, e^{\left((u -
\eta^+)/\sqrt{2}\right)(-\sqrt{2}\eta^-)} \nonumber\\
&=& 2^{(n+m)/2}\, e^{\eta^2}\, \frac{n!}{2\pi i} \oint \rmd u\,
\frac{u^m}{\left(u -
\eta^+\right)^{n+1}}\, e^{-u \eta^-} \nonumber\\
&=& 2^{(n+m)/2}\, e^{\eta^2}\, \left.\frac{\rmd^n}{{\rmd u}^n}\,
u^m\, e^{-u\eta^-} \right|_{u =
\eta^+} \nonumber\\
&=& 2^{(n+m)/2}\, e^{\eta^2}\, (\eta^-)^{n-m} \frac{\rmd^n}{{\rmd
(\eta^2)}^n}\, (\eta^2)^m\, e^{-\eta^2}
\nonumber\\
&=& n!\,2^{(n+m)/2}\, (\eta^{+})^{(m-n)}\,
L^{m-n}_n(\eta^2)\nonumber
\end{eqnarray}
where we have used the Rodrigues' representation for Laguerre
polynomials \cite[Eq.~(13.47)]{Arfken:1995a}
\begin{equation}\label{apInmeq:13}
L^{k}_n(x) =  \frac{e^x\, x^{-k}}{n!}\, \frac{\rmd^n}{\rmd x^n}\,
x^{n+k}\, e^{-x}\,, \qquad n,k \geq 0\,.
\end{equation}
With the result from Eq.~(\ref{apInmeq:12}), we can now express the
$I_{n,m}$ as
\begin{equation}\label{apInmeq:14}
I_{n,m} = \sqrt{\frac{n!}{m!}}\ e^{-\eta^2/2}\, e^{i \phi_0}\,
\left(\eta_x + i\, \eta_y\right)^{m-n}\, L_n^{m-n}(\eta^2)\,,\quad m
\geq n \geq 0 \,,
\end{equation}

For the case when $n > m$ we first integrate over $t$ in
Eq.~(\ref{apInmeq:8}) and follow a similar procedure to find
\begin{equation}\label{apInmeq:15}
I_{n,m} = \sqrt{\frac{m!}{n!}}\ e^{-\eta^2/2}\, e^{i \phi_0}\,
\left(-\eta_x + i\, \eta_y\right)^{n-m}\, L_m^{n-m}(\eta^2)\,,\quad
n \geq m \geq 0 \,.
\end{equation}

\section{Result of trace\label{ap:trace}}
We can express the trace result for the average of the squared
scattering amplitude from from Eq.~(\ref{eq:M2avgc}) as a sum of
terms
\begin{equation}\label{aptraceeq:1}
\overline{\left| \mathcal{M} \right|^2} = \frac{G_F^2}{2^{9} E E'
E_{n_1} E_{n_2}}\ \sum_{i=1}^{16} A_i\,  \mathrm{T}_i\,,
\end{equation}
where the coefficients $A_i$ depend on the products of $I_{n,m}$ and
$I^\ast_{n',m'}$ defined in Eq.~(\ref{eq:Inma}) and presented in
appendix~\ref{ap:Inm}, and the $T_i$ are the parts that depend on
the contraction of the traces in Eq.~(\ref{eq:M2avgc}).  The results
are as follows:
\begin{eqnarray}
A_1 &=& I_{n_1, n_2} I^\ast_{n_1, n_2} \\
T_1 &=& \mathrm{Tr} \Bigl\{ \gamma^\sigma \left(G_V^\pm -
\gamma^5\right) \left[ m (1 - \sigma^3) + \notp{p}_{1 \parallel} +
\notp{q}_{1 \parallel} \gamma^5\right] \gamma^\mu \left(G_V^\pm -
\gamma^5\right) \left[- m (1 - \sigma^3) + \notp{p}_{2
\parallel} + \notp{q}_{2 \parallel} \gamma^5\right] \Bigr\}
\nonumber\\
T_1 &=& -2^7 \left({G_V^\pm}^2 - 1\right) m_e^2 \left(p_x p_x' + p_y
p_y'\right) + 2^6 \left(G_V^\pm + 1 \right)^2 (E - p_z) (E' - p_z')
(E_{n_1} + p_{1\,z}) (E_{n_2} + p_{2\,z}) \nonumber\\
&& + 2^6 \left(G_V^\pm - 1 \right)^2 (E + p_z) (E' + p_z') (E_{n_1}
- p_{1\,z}) (E_{n_2} - p_{2\,z})
\end{eqnarray}

\begin{eqnarray}
A_2 &=& I_{n_1, n_2-1} I^\ast_{n_1, n_2}\\
T_2 &=& \mathrm{Tr} \Bigl\{ \gamma^\sigma \left(G_V^\pm -
\gamma^5\right) \left[ m (1 - \sigma^3) + \notp{p}_{1 \parallel} +
\notp{q}_{1 \parallel} \gamma^5\right] \gamma^\mu \left(G_V^\pm -
\gamma^5\right) (-\sqrt{2n_2 eB}) \left(\gamma^1 +
i \gamma^2\right) \Bigr\} \nonumber\\
T_2 &=& - 2^6 \left(G_V^\pm + 1 \right)^2 \sqrt{2n_2 eB}\,%
(p_x + i\, p_y) (E' - p_z') (E_{n_1} + p_{1\,z})\nonumber\\
&& - 2^6 \left(G_V^\pm - 1 \right)^2 \sqrt{2n_2 eB}\,%
(p_x' + i\, p_y') (E + p_z) (E_{n_1} - p_{1\,z})
\end{eqnarray}

\begin{eqnarray}
A_3 &=& I_{n_1, n_2} I^\ast_{n_1, n_2-1}\ =\ A^\ast_2 \\
T_3 &=& T^\ast_2
\end{eqnarray}

\begin{eqnarray}
A_4 &=& I_{n_1, n_2-1} I^\ast_{n_1, n_2-1} \\
T_4 &=& \mathrm{Tr} \Bigl\{ \gamma^\sigma \left(G_V^\pm -
\gamma^5\right) \left[ m (1 - \sigma^3) + \notp{p}_{1 \parallel} +
\notp{q}_{1 \parallel} \gamma^5\right]  \gamma^\mu \left(G_V^\pm -
\gamma^5\right) \left[- m (1 + \sigma^3) + \notp{p}_{2
\parallel} - \notp{q}_{2 \parallel} \gamma^5\right] \Bigr\}
\nonumber\\
T_4 &=& 2^7 \left({G_V^\pm}^2 - 1\right) m_e^2 (E + p_z) (E' - p_z')
+ 2^6 \left(G_V^\pm + 1 \right)^2 (E + p_z) (E' - p_z')
(E_{n_1} + p_{1\,z}) (E_{n_2} - p_{2\,z}) \nonumber\\
&& + 2^6 \left(G_V^\pm - 1 \right)^2 (E + p_z) (E' - p_z') (E_{n_1}
- p_{1\,z}) (E_{n_2} + p_{2\,z})
\end{eqnarray}

\begin{eqnarray}
A_5 &=& I_{n_1, n_2} I^\ast_{n_1-1, n_2} \\
T_5 &=& \mathrm{Tr} \Bigl\{ \gamma^\sigma \left(G_V^\pm -
\gamma^5\right) (\sqrt{2n_1 eB}) \left(\gamma^1 + i \gamma^2\right)
\gamma^\mu \left(G_V^\pm - \gamma^5\right) \left[- m (1 - \sigma^3)
+ \notp{p}_{2
\parallel} + \notp{q}_{2 \parallel} \gamma^5\right] \Bigr\}
\nonumber\\
T_5 &=& 2^6 \left(G_V^\pm + 1 \right)^2 \sqrt{2n_1 eB}\,%
(p_x' + i\, p_y') (E - p_z) (E_{n_2} + p_{2\,z}) \nonumber\\
&& + 2^6 \left(G_V^\pm - 1 \right)^2 \sqrt{2n_1 eB}\,%
(p_x + i\, p_y) (E' + p_z') (E_{n_2} - p_{2\,z})
\end{eqnarray}

\begin{eqnarray}
A_6 &=& I_{n_1, n_2-1} I^\ast_{n_1-1, n_2} \\
T_6 &=& \mathrm{Tr} \Bigl\{ \gamma^\sigma \left(G_V^\pm -
\gamma^5\right) (\sqrt{2n_1 eB}) \left(\gamma^1 + i \gamma^2\right)
\gamma^\mu \left(G_V^\pm - \gamma^5\right) (-\sqrt{2n_2 eB})
\left(\gamma^1 +
i \gamma^2\right) \Bigr\} \nonumber\\
T_6 &=& -2^6 \left(G_V^\pm + 1 \right)^2 \sqrt{2n_1 eB}\,%
\sqrt{2n_2 eB}\, (p_x + i\, p_y) (p_x' + i\, p_y') \nonumber\\
&& -2^6 \left(G_V^\pm - 1 \right)^2 \sqrt{2n_1 eB}\,%
\sqrt{2n_2 eB}\, (p_x + i\, p_y) (p_x' + i\, p_y')
\end{eqnarray}

\begin{eqnarray}
A_7 &=& I_{n_1, n_2} I^\ast_{n_1-1, n_2-1} \\
T_7 &=& \mathrm{Tr} \Bigl\{ \gamma^\sigma \left(G_V^\pm -
\gamma^5\right) (\sqrt{2n_1 eB}) \left(\gamma^1 + i \gamma^2\right)
\gamma^\mu \left(G_V^\pm - \gamma^5\right) (-\sqrt{2n_2 eB})
\left(\gamma^1 -
i \gamma^2\right) \Bigr\} \nonumber\\
T_7 &=& -2^6 \left(G_V^\pm + 1 \right)^2 \sqrt{2n_1 eB}\,%
\sqrt{2n_2 eB}\, (p_x - i\, p_y) (p_x' + i\, p_y') \nonumber\\
&& -2^6 \left(G_V^\pm - 1 \right)^2 \sqrt{2n_1 eB}\,%
\sqrt{2n_2 eB}\, (p_x + i\, p_y) (p_x' - i\, p_y')
\end{eqnarray}

\begin{eqnarray}
A_8 &=& I_{n_1, n_2-1} I^\ast_{n_1-1, n_2-1} \\
T_8 &=& \mathrm{Tr} \Bigl\{ \gamma^\sigma \left(G_V^\pm -
\gamma^5\right) (\sqrt{2n_1 eB}) \left(\gamma^1 + i \gamma^2\right)
\gamma^\mu \left(G_V^\pm - \gamma^5\right) \left[- m (1 + \sigma^3)
+ \notp{p}_{2
\parallel} - \notp{q}_{2 \parallel} \gamma^5\right] \Bigr\}
\nonumber\\
T_8 &=& 2^6 \left(G_V^\pm + 1 \right)^2 \sqrt{2n_1 eB}\,%
(p_x' + i\, p_y') (E + p_z) (E_{n_2} - p_{2\,z})\nonumber\\
&& + 2^6 \left(G_V^\pm - 1 \right)^2 \sqrt{2n_1 eB}\,%
(p_x + i\, p_y) (E' + p_z') (E_{n_2} + p_{2\,z})
\end{eqnarray}

\begin{eqnarray}
A_9 &=& I_{n_1-1, n_2} I^\ast_{n_1, n_2}\ =\ A_5^\ast \\
T_9 &=& T_5^\ast
\end{eqnarray}

\begin{eqnarray}
A_{10} &=& I_{n_1-1, n_2-1} I^\ast_{n_1, n_2}\ =\ A_7 \\
T_{10} &=& T_7^\ast
\end{eqnarray}

\begin{eqnarray}
A_{11} &=& I_{n_1, n_2} I^\ast_{n_1, n_2-1}\ =\ A^\ast_6 \\
T_{11} &=& T^\ast_6
\end{eqnarray}

\begin{eqnarray}
A_{12} &=& I_{n_1-1, n_2-1} I^\ast_{n_1, n_2-1}\ =\ A_8^\ast \\
T_{12} &=& T_8^\ast
\end{eqnarray}

\begin{eqnarray}
A_{13} &=& I_{n_1-1, n_2} I^\ast_{n_1-1, n_2} \\
T_{13} &=& \mathrm{Tr} \Bigl\{ \gamma^\sigma \left(G_V^\pm -
\gamma^5\right) \left[ m (1 + \sigma^3) + \notp{p}_{1 \parallel} -
\notp{q}_{1 \parallel} \gamma^5\right] \gamma^\mu \left(G_V^\pm -
\gamma^5\right) \left[- m (1 - \sigma^3) + \notp{p}_{2
\parallel} + \notp{q}_{2 \parallel} \gamma^5\right] \Bigr\}
\nonumber\\
T_{13} &=& 2^7 \left({G_V^\pm}^2 - 1\right) m_e^2 (E - p_z) (E' +
p_z') + 2^6 \left(G_V^\pm + 1 \right)^2 (E - p_z) (E' + p_z')
(E_{n_1} - p_{1\,z}) (E_{n_2} + p_{2\,z}) \nonumber\\
&& + 2^6 \left(G_V^\pm - 1 \right)^2 (E - p_z) (E' + p_z') (E_{n_1}
+ p_{1\,z}) (E_{n_2} - p_{2\,z})
\end{eqnarray}

\begin{eqnarray}
A_{14} &=& I_{n_1-1, n_2-1} I^\ast_{n_1-1, n_2} \\
T_{14} &=& \mathrm{Tr} \Bigl\{ \gamma^\sigma \left(G_V^\pm -
\gamma^5\right) \left[ m (1 + \sigma^3) + \notp{p}_{1 \parallel} -
\notp{q}_{1 \parallel} \gamma^5\right] \gamma^\mu \left(G_V^\pm -
\gamma^5\right) (-\sqrt{2n_2 eB}) \left(\gamma^1 + i
\gamma^2\right) \Bigr\} \nonumber\\
T_{14} &=& - 2^6 \left(G_V^\pm + 1 \right)^2 (\sqrt{2n_2 eB}) (p_x
+ i p_y) (E' + p_z') (E_{n_1} - p_{1\,z}) \nonumber\\
&& - 2^6 \left(G_V^\pm - 1 \right)^2 (\sqrt{2n_2 eB}) (p_x' + i
p_y') (E - p_z) (E_{n_1} + p_{1\,z})
\end{eqnarray}

\begin{eqnarray}
A_{15} &=& I_{n_1-1, n_2} I^\ast_{n_1-1, n_2-1}\ =\ A^\ast_{14} \\
T_{15} &=& T_{14}^\ast
\end{eqnarray}

\begin{eqnarray}
A_{16} &=& I_{n_1-1, n_2-1} I^\ast_{n_1-1, n_2-1} \\
T_{16} &=& \mathrm{Tr} \Bigl\{ \gamma^\sigma \left(G_V^\pm -
\gamma^5\right) \left[ m (1 + \sigma^3) + \notp{p}_{1 \parallel} -
\notp{q}_{1 \parallel} \gamma^5\right] \gamma^\mu \left(G_V^\pm -
\gamma^5\right) \left[- m (1 + \sigma^3) + \notp{p}_{2
\parallel} - \notp{q}_{2 \parallel} \gamma^5\right] \Bigr\}
\nonumber\\
T_{16} &=& - 2^7 \left({G_V^\pm}^2 - 1\right) m_e^2 (p_x p_x' + p_y
p_y') + 2^6 \left(G_V^\pm + 1 \right)^2 (E + p_z) (E' + p_z')
(E_{n_1} - p_{1\,z}) (E_{n_2} - p_{2\,z}) \nonumber\\
&& + 2^6 \left(G_V^\pm - 1 \right)^2 (E - p_z) (E' - p_z') (E_{n_1}
+ p_{1\,z}) (E_{n_2} + p_{2\,z})\,.\quad
\end{eqnarray}

\end{document}